\documentclass[12pt]{article}

\pdfoutput=1

\usepackage{color}

\usepackage{amsmath,amssymb,latexsym}

\usepackage{cite}

\usepackage{fouriernc}

\usepackage{caption}

\usepackage{fdsymbol}

\usepackage{indentfirst}

\let\a=\alpha \let\b=\beta \let\g=\gamma \let\d=\delta \let\e=\epsilon
\let\i=\iota \let\k=\kappa
\let\l=\lambda \let\m=\mu \let\n=\nu \let\x=\xi \let\p=\pi 
\let\s=\sigma 

       \let\D=\Delta \let\Th=\Theta \let\L=\Lambda
\let\X=\Xi  \let\S=\Sigma  \let\Y=\Psi
 
\let\la=\label  
  
 \def\bd{\begin{document}} \def\ed{\end{document}}
\def\ds{\documentstyle} \let\fr=\frac \let\bl=\bigl \let\br=\bigr
\let\Br=\Bigr \let\Bl=\Bigl
\let\bm=\bibitem
\let\na=\nabla
\def\tU{{\widetilde U}}
\let\pa=\partial \let\ov=\overline
\def\ie{{\it i.e.\ }}
\newcommand{\be}{\begin{equation}}
\newcommand{\ee}{\end{equation}}
\def\ba{\begin{array}}
\def\ea{\end{array}}

\def\bei{\begin{itemize}}
\def\eei{\end{itemize}}

\def\ben{\begin{enumerate}}
\def\een{\end{enumerate}}

\def\ft#1#2{{\textstyle{{\scriptstyle #1}\over {\scriptstyle #2}}}}
\def\fft#1#2{{#1 \over #2}}
\def\F#1#2{{ F_{#1}^{(#2)} }}
\def\cF#1#2{{ {\cal F}_{#1}^{(#2)} }}

\def\R{{\bf R}}
\def\sst#1{{\scriptscriptstyle #1}}
\def\oneone{\rlap 1\mkern4mu{\rm l}}
\def\e7{E_{7(+7)}}
\def\td{\tilde}
\def\wtd{\widetilde}
\def\im{{\rm i}}
\def\bog{Bogomol'nyi\ }
\newcommand{\ho}[1]{$\, ^{#1}$}
\newcommand{\hoch}[1]{$\, ^{#1}$}
\newcommand{\bea}{\begin{eqnarray}}
\newcommand{\eea}{\end{eqnarray}}
\newcommand{\ra}{\rightarrow}
\newcommand{\lra}{\longrightarrow}
\newcommand{\Lra}{\Leftrightarrow}
\newcommand{\ap}{\alpha^\prime}
\newcommand{\bp}{\tilde \beta^\prime}
\newcommand{\cB}{{\cal B}}
\newcommand{\cO}{{\cal O}}
\newcommand{\vecx}{\vec{x}}
\newcommand{\vecy}{\vec{y}}
\newcommand{\vecp}{\vec{p}}
\newcommand{\vecq}{\vec{q}}
\newcommand{\tr}{{\rm tr} }
\newcommand{\Tr}{{\rm Tr} }
\newcommand{\NP}{Nucl. Phys. }

\newcommand{\cL}{{\cal L}}
\newcommand{\cA}{{\cal A}}
\newcommand{\cT}{{\cal T}}
\newcommand{\cD}{{\cal D}}
\newcommand{\cH}{{\cal H}}

\def\th{\theta}

\def\sst#1{{\scriptscriptstyle #1}}
\def\0{{\sst{(0)}}}
\def\1{{\sst{(1)}}}
\def\2{{\sst{(2)}}}
\def\3{{\sst{(3)}}}
\def\4{{\sst{(4)}}}
\def\5{{\sst{(5)}}}
\def\6{{\sst{(6)}}}
\def\7{{\sst{(7)}}}
\def\8{{\sst{(8)}}}
\def\9{{\sst{(9)}}}
\def\p{{\sst{(p)}}}
\def\q{{\sst{(q)}}}

\def\ssa{{\sst{(\alpha)}}}
\def\ssb{{\sst{(\beta)}}}
\def\ssg{{\sst{(\gamma)}}}
\def\j{{\sst{(j)}}}

\def\ve{\varepsilon}
\def\vf{\varphi}
\def\F{\Phi}
\def\wg{\wedge}

\def\thb{\bar{\theta}}
\def\Thb{\bar{\Theta}}
\def\barp{\bar{p}}
\def\barq{\bar{q}}
\def\barc{\bar{c}}
\def\bard{\bar{d}}
\def\e{\epsilon}

\def \bi{\bibitem}
\def \la {\label}

\def \l {\lambda}
\def\foot{\footnote}
\def \tl  {{\tilde \l}}
\def \sql {{\sqrt \l}}
\def \adss {$AdS_5 \times S^5$\ }
\newcommand{\rf}[1]{(\ref{#1})}
\def \ov {\over}

\def\Th{\Theta}
\def\vth{\vartheta}
\def\btheta{{\bar\theta}}
\def\ttheta{{{\tilde\theta}}}
\def\bttheta{{{\bar\ttheta}}}
\def\vth{\vartheta}

\def\ra{\rightarrow}
\def\N{{\cal N}}
\def\uM{\underline{M}}
\def\uA{\underline{A}}
\def\uN{\underline{N}}
\def\uP{\underline{P}}
\def\ua{\underline{a}}
\def\ub{\underline{b}}
\def\uc{\underline{c}}
\def\ud{\underline{d}}
\def\ue{\underline{e}}
\def\uf{\underline{f}}
\def\ui{\underline{i}}
\def\uj{\underline{j}}
\def\uk{\underline{k}}
\def\ual{\underline{\alpha}}
\def\ube{\underline{\beta}}
\def\um{\underline{m}}
\def\un{\underline{n}}
\def\up{\underline{p}}
\def\uq{\underline{q}}
\def\ur{\underline{r}}
\def\us{\underline{s}}

\def\umu{\underline{\mu}}
\def\unu{\underline{\nu}}
\def\ula{\underline{\l}}
\def\uka{\underline{\k}}
\def\usi{\underline{\s}}
\def\urh{\underline{\r}}
\def\cc{\circ}
\def\eqv{\equiv}

\def\ni{\noindent}

\def\Ep{E^{{}^{(+)}}}
\def\Em{E^{{}^{(-)}}}

\def\Mp{M^{{}^{(+)}}}
\def\Mm{M^{{}^{(-)}}}

\def \ha{{1\ov 2}}

\def\r{\rho}

\def\Y{{\rm Y}}
\def\X{{\rm X}}
\def\tY{\tilde{\rm Y}}
\def\tX{\tilde{\rm X}}
\def\dY{\dot{\rm Y}}
\def\dX{\dot{\rm X}}

\def \J {\mathcal{J}}
\def \del {\partial}

\def\dF{\dot{F}}
\def\dG{\dot{G}}
\def\df{\dot{f}}
\def\dx{\dot{x}}
\def \E {{\cal E}}
\def \S {{\cal S}}
\def \J {{\cal J}}

\def\ms{\mathcal{S}}
\def\mj{\mathcal{J}}
\def\soj{\fr{\ms}{\mj}}
\def \R {{\bf R}}
\def \om {\omega}
\def \bE {\bar E}
\def \x {{\cal X}}

\def \bi{\bibitem}
\def \la {\label}

\def \l {\lambda}
\def\foot{\footnote}
\def \tl  {{\tilde \l}}
\def \sql {{\sqrt \l}}
\def \adss {$AdS_5 \times S^5$\ }
\def \ov {\over}

\def \varpi {{\rm w}}

\def\thb{\bar{\theta}}
\def\Thb{\bar{\Theta}}
\def\mb{\bar{\m}}
\def\ab{\bar{\a}}
\def\zb{\bar{z}}
\def\psib{\bar{\psi}}
\def\barp{\bar{p}}
\def\barq{\bar{q}}
\def\barc{\bar{c}}
\def\bard{\bar{d}}
\def\e{\epsilon}
\def\wb{\bar{w}}
\def\lb{\bar{\l}}
\def\Jb{\bar{J}}
\def\Nb{\bar{N}}
\def\Zb{\bar{Z}}
\def\pab{\bar{\pa}}

\def\At{\tilde{A}}
\def\Bt{\tilde{B}}
\def\Ct{\tilde{C}}
\def\Dt{\tilde{D}}
\def\Et{\tilde{E}}
\def\Ft{\tilde{F}}
\def\Gt{\tilde{G}}
\def\Ht{\tilde{H}}
\def\Mt{\tilde{M}}
\def\Rt{\tilde{R}}
\def\at{\tilde{a}}
\def\bt{\tilde{b}}
\def\ct{\tilde{c}}
\def\dt{\tilde{d}}
\def\et{\tilde{e}}
\def\ft{\tilde{f}}
\def\gt{\tilde{g}}
\def\mt{\tilde{\mu}}
\def\nt{\tilde{\nu}}

\def\asth{\hat{*}}
\def\phh{\hat{\phi}}

\def\bA{{\bf A}}

\def\ola{\overleftarrow}
\def\ora{\overrightarrow}
\def\alt{\tilde{\a}}

\def\eh{\hat{e}}
\def\eph{\hat{\e}}
\def\ph{\hat{p}}
\def\alh{\hat{\a}}
\def\beh{\hat{\b}}
\def\gah{\hat{\g}}
\def\Fh{\hat{F}}
\def\muh{\hat{\m}}
\def\nuh{\hat{\n}}
\def\thh{\hat{\th}}
\def\dh{\hat{d}}
\def\ih{\hat{i}}
\def\jh{\hat{j}}
\def\kh{\hat{k}}
\def\deh{\hat{\d}}
\def\wh{\hat{w}}
\def\lah{\hat{\l}}
\def\Ah{\hat{A}}
\def\Ch{\hat{C}}
\def\Omh{\hat{\Omega}}

\def\xh{\hat{x}}

\def\ps{\rlap{\, /}\;\,p }
\def\ks{\rlap{\, /}\;\,k }

\def\gym{g_{YM}}

\def\adot{\dot{a}}
\def\bdot{\dot{b}}
\def\bpa{\bar{\pa}}

\def\pr{\prime}
\def\ssk{\medskip}
\def\bsk{\bigskip}

\def\clb{\color{blue}}
\def\clr{\color{red}}
\def\clv{\color{violet}}

\def\t{\tau}
\def\cM{\mathcal{M}}
\def\S{\Sigma}

\def\N{\nabla}

\def\cR{\mathcal{R}}
\def\cL{\mathcal{L}}

\def\hb{\hbar}
\def\an{\hat{a}}
\def\ac{\hat{a}^\dag}
\def\hp{\hat{p}}

\def\Ec{{\cal E}}

\DeclareMathOperator{\arccosh}{arccosh}

\begin{document}

\title{
{\Large{\bf 
Studying KARMA\\
(K{\small err} A{\small nd} R{\small elativistic} M{\small atter} A{\small round})
}}
}
{
\author{
A.J. Nurmagambetov$\,^{\spadesuit,\vardiamondsuit,\varheartsuit}$\footnote{ajn@kipt.kharkov.ua; a.j.nurmagambetov@gmail.com}
\\ \\
$\,^{\spadesuit}${ \it {\normalsize Akhiezer Institute for Theoretical Physics of NSC KIPT}}
\\
{ \it {\normalsize 1 Akademichna St., Kharkiv 61108, Ukraine} }
\\
$\,^{\vardiamondsuit}${ \it {\normalsize Department of Physics \& Technology, Karazin Kharkiv National University,}}
\\
{ \it {\normalsize 4 Svobody Sq., Kharkiv 61022, Ukraine} }
\\
$\,^{\varheartsuit}${ \it {\normalsize Usikov Institute of Radiophysics and Electronics}}\\
{ \it {\normalsize 12 Ak. Proskury, Kharkiv 61085, Ukraine} }
}

\date{}

\maketitle

\abstract{
We construct series of solutions for the Kerr-type rotating black hole with non-trivial matter in flat and (A)dS backgrounds. Symmetry arguments and singularity analysis in the proposed black hole models fix the free parameters of the solutions, and the study of popular energy conditions makes it possible to impose constraints on configuration and field content of external matter. The resulted geometry of spacetimes is featured by a special type singularity in the north and south pole directions, inducing the bipolar outflow of particles from black holes. As a step toward the construction of a non-stationary rotating black hole solution in the presence of matter, we explore the zero angular momentum limit of the constructed metrics. The use of the Eddington-Finkelstein coordinates allows us to find a generalization of the proposed construction to the Vaidya-type black hole. Finally, employing the Newman-Janis algorithm, we find the corresponding generalization of the Kerr-Vaidya black hole solution.

}

\bsk\bsk
PACS: 04.70.-s, 04.70.Bw, 04.20.Jb

\ssk
\noindent {\it Keywords:} Kerr-Vaidya black holes, Newman-Janis algorithm, energy conditions

\newpage



\section{Introduction}

Kerr's discovery of a new type of solutions of Einstein's vacuum equations had initiated not only a new direction in the study of black holes with properties of compact astrophysical objects existing in nature, but also raised a number of fundamental questions about the mechanisms of formation of rotating black holes. The central question about the sources of the final in its evolution stationary configuration of a rotating black hole without matter fields was raised by Kerr himself in his pioneering paper \cite{Kerr:1963ud}, and the pool of open questions accumulated for a decade and a half after the publication of Kerr's paper was reduced by Krasinski \cite{Krasinski:1976vyc} to four main points. They are (cf. Ref. \cite{Krasinski:1976vyc}): ``i) Identifying possible sources by investigation of singularities and physical interpretation of parameters of the Kerr metric; ii) Eliminating some types of sources due to contradictions or inconsistencies to which they lead; iii) Constructing physically acceptable configurations matched to the Kerr metric only approximately; iv) Constructing odd configurations with unphysical properties matched to the Kerr metric exactly.'' The main motivation for this paper is the first two points noted by Krasinski. Although partially the work also addresses the problems identified in the remaining two points.

The construction of the metric of a rotating black hole is traditionally based on the use of the seed metric for a static black hole with the subsequent application of either the standard Newman-Janis algorithm \cite{Newman:1965tw} or its generalizations \cite{Drake:1998gf,Keane:2014sta,Erbin:2014aya} and modifications \cite{Azreg-Ainou:2014pra}. A less traditional way is to use ellipsoidal coordinates, which are more adapted to describe a freely rotating elastic body. Note that the latter approach has been used in a number of works oriented at the attempt to construct more general Kerr-type solutions, that could shed light on the ways of resolving the problems mentioned by Krasinski. Here, in addition to the abovementioned Krasinski's paper, we can mention the work of Nikolic and Pantic \cite{Nikolic:2012vr}, which analyzed possible extensions of the Kerr metric, and conditions on the free parameters eventually leading  to the Kerr geometry. It is also noteworthy the work of Dadhich \cite{Dadhich:2013qx}, which justifies the Kerr solution as the most natural one for an ellipsoidal coordinate system. The present work intensively uses the approach of Dadhich, as well as symmetry properties of the Kerr solution underlying the choice of generalized solutions for rotating black holes with a non-trivial stress-energy tensor of matter. With respect to the symmetry properties of an axisymmetric solution to the Einstein equations our approach is ideologically close to the approach of \cite{Konoplya:2016jvv}. However, we specify the unspecified functions in the solution of \cite{Konoplya:2016jvv} using different energy conditions on the stress-energy tensor of matter. Some progress for the Kerr-type solution with matter fields has been achieved in Refs. \cite{Kim:2019hfp,Nurmagambetov:2020ann}.

The paper is organized as follows. In Section 2 we set up the stage for the construction of the generalized Kerr-type solution. Here we outline a way to introduce the deformation function into the metric of a rotating black hole in ellipsoidal coordinates with symmetry requirements. Then we focus on the overall structure of the effective stress-energy tensor of matter generating by the deformation function and restrict the consideration with the stress-energy tensor of an anisotropic fluid. The latter allows us to restrict the deformation function to four free parameters. These parameters can be determined both from the requirement to have as far as possible the same singularities of the metric as the standard Kerr, and from the different energy conditions on the matter fields to be physical. As a result the constructed solution possesses, besides standard points of singularity of the Kerr black hole, a special type singularity in directions of the north and south poles, which is responsible for occurrence of jets of outcoming particles.  We close this section with constructing the (A)dS generalization of the obtained solutions and discussing its zero cosmological constant limit.  

In Section 3 we study the non-rotating limit of the obtained generalization of the Kerr solution in different coordinates. We analyse the Schwarzschild-type solution in the presence of matter, then we do the same for a generalized static black hole in the null coordinates. After that, we construct the Vaidya-type solution and make the analysis of the energy conditions in this case. Here we come to a curious conclusion about the matter content of the accretion disk: in the null (Eddington-Finkelstein) coordinates the stress-energy tensor of matter becomes free from such pathologies as tachyonic-type matter or energy negativity if the conditionally stable closed trajectories of particles in the equatorial plane correspond solely to massive particles. In Section 4 we extend the construction to the case of Kerr-Vaidya-type black hole and end up with general form of the metric for a rotating non-stationary black hole with matter.

The last section contains the discussion of the results and our conclusions. Technical details of computations are collected in four Appendices. In our notation the black hole mass is renormalized by the Newton constant, i.e., $M=m_0 G$, where $m_0$ is the actual mass of the black hole.

\section{Kerr type solution with matter fields: metrics, singularities and energy conditions}

In this part of the paper we set up the stage for the construction of generalized Kerr-type solution by use of the ellipsoidal coordinate system and the symmetry arguments. Having obtained a general expression for the metric, we investigate its singularities and constraints on non-fixed parameters arising from the requirement that matter fields be physically admissible.

\subsection{From a Deformed Ellipsoid to a Deformed Kerr}

Let us briefly recap the main steps in the constructions of \cite{Dadhich:2013qx}, which is based on the previous consideration of \cite{Krasinski:1976vyc,Nikolic:2012vr}. The starting point is the line element of 3D Euclidean space in the ellipsoidal coordinates, which is elementary extended to flat Minkowski spacetime:
\be
ds^2_{4D}=-dt^2+\fr{\r^2}{r^2+a^2}dr^2+\r^2 d\th^2+(r^2+a^2)\sin^2\th d\vf^2,
\qquad \r^2=r^2+a^2 \cos^2\th .
\la{Minkellips}
\ee
Alternately, we can rewrite \rf{Minkellips} as
\be
ds^2_{4D}=-A\left(dt-a\sin^2 \th \,d\vf\right)^2+\fr1{A} dr^2+\r^2 d\th^2+\fr{\sin^2\th}{\r^2}\left[(r^2+a^2) d\vf -a dt \right]^2 ,
\la{Rotellips}
\ee
so that to match eqs. \rf{Rotellips} and \rf{Minkellips} one has to choose 
\[
A=\fr{r^2+a^2}{\r^2} .
\]
Then, as it was indicated in Ref. \cite{Dadhich:2013qx}, the interval in \rf{Rotellips} is nothing but the Kerr metric for a BH of zero mass. Replacing $A$ factor with
\be
A(r,\th)=\fr{\D}{\r^2},\quad \D=r^2+a^2-2Mr,
\la{AdefKerr}
\ee
results in the standard Kerr metric in the Boyer-Lindquist coordinates.

Now, we will enhance the construction to include external matter. The requirement of having the symmetry under $\vf \ra -\vf$ (see Ref. \cite{Krasinski:1976vyc} for details) is realized in the following deformation of 3D Euclidean space,
\be
ds^2_{3D}=f^2(r,\th) dr^2+\r^2 d\th^2+(r^2+a^2)\sin^2\th d\vf^2,\quad \r^2=r^2+a^2 \cos^2\th,
\la{ellipsDef}
\ee
which is not utmostly general. We propose to generalize this metric to
\be
ds^2_{3D}=e^{2W(r,\th)} dr^2+e^{2W(r,\th)}\r^2 d\th^2+(r^2+a^2)\sin^2\th d\vf^2,
\la{ellipsDefm}
\ee
that also keeps the symmetry under the $\vf$ inversion. Then, employing the formalism of \cite{Dadhich:2013qx}, we arrive at the following generalization of the metric \rf{Rotellips}:
\be
ds^2=-\fr{\D}{\r^2}\left(dt-a\sin^2 \th \,d\vf\right)^2+\fr{\tilde{\r}^2}{\D} dr^2+\tilde{\r}^2 d\th^2+\fr{\sin^2 \th}{\r^2}\left[\left(r^2+a^2\right)d\vf-a dt \right]^2.
\la{Rotellipsm}
\ee
Here, as in eq. \rf{Rotellips}, 
\be
\D=r^2+a^2-2Mr,\quad \r^2=r^2+a^2 \cos^2 \th,
\la{Drho}
\ee
and
\be
\tilde{\r}^2=e^{2W(r,\th)}\r^2.
\la{rhotil}
\ee
Having fixed the spacetime geometry with the line element \rf{Rotellipsm}, we proceed to the study of the properties of matter compatible with such geometry.

\subsection{The stress-energy tensor: general consideration and specifics}

Direct calculations of the Einstein tensor $G_{mn}\equiv R_{mn}-1/2 g_{mn} R$ for the metric \rf{Rotellipsm} with a non-trivial deformation function $W(r,\th)$ lead to the conclusion on the mandatory presence of matter fields: 
\be
G_{mn}=8\pi G T_{mn}.
\la{Eins}
\ee
Any stress-energy tensor $T_{mn}$ of any type of relativistic matter
can be equivalently presented in the form of (see Section 7.8 of \cite{Alcubierre:2008}) 
\be
T_{mn}=\left({\cal E}+P\right)u_m u_n+P g_{mn}+q_m u_n+q_n u_m+\t_{mn},
\la{Tmndef}
\ee
where, in the nomenclature of a viscous fluid, ${\cal E}$ and $P$ are the energy and the pressure density, respectively; $q_m$ is the momentum (heat) flux vector, and $\t_{mn}$ is the shear tensor. For the metric determined by \rf{Rotellipsm}, we can directly use the expression for the 4-velocity vector $u_m$ from \cite{Nurmagambetov:2022wcw}:
\be
\begin{split}
&u_m=\left(\sqrt{1-\fr{2Mr}{\r^2}},0,0,\fr{2aMr\sin^2 \th}{\r^2\sqrt{1-\fr{2Mr}{\r^2}}}\right),\qquad u^m u_m=-1.
\end{split}
\la{udef}
\ee
However, there is a subtlety related to the stress-energy tensor \rf{Tmndef}: it is silently supposed that the pressure density is isotropic, that is, of the same value $P$ in all directions. Due to the structure of the considered metric \rf{Rotellipsm}, the isotropy is lost, so that the close inspection of non-trivial components of the Einstein tensor results in the following stress-energy tensor of an anisotropic fluid   
\be
T_{mn}=\left({\cal E}+P\right)u_m u_n+P g_{mn}+\left(P_\th-P\right)g_{\th\th}+\left(P_\vf-P\right)g_{\vf\vf} .
\la{Tmnan}
\ee
Here $P$ is the radial pressure, and $P_\th$, $P_\vf$ are tangential pressures.
Explicitly, 
\be
\begin{split}
{\cal E}=-\fr{e^{-2W}}{8\pi G \r^2}\left(\D \pa^2_r W+\left(r-M\right)\pa_r W+\pa^2_\th W \right),\\ 
P=-P_\th=\fr{e^{-2W}}{8\pi G \r^2} \left[\left(r-M \right)\pa_r W-\cot \th \pa_\th W \right],
\end{split}
\la{Edef}
\ee
\be
\begin{split}
P_\vf=&\fr{e^{-2W}}{8\pi G \r^2}\Big[\left(r-M \right)\pa_r W+\fr1{\D \r^4-4M^2 a^2 r^2 \sin^2\th} \times \\
\times & \left(\D \r^4 \left(\D \pa^2_r W+\pa^2_\th W \right)+2\pa_\th W M^2 a^2 r^2 \sin 2\th \right) \Big].
\end{split}
\la{Pvfdef}
\ee
This simple representation of $T_{mn}$ is not common; it becomes possible upon demanding $G_{r\th}=0$, which constraints the still unconstrained deformation function $W(r,\th)$. We now turn to the analysis of possible solutions to this equation.

\subsection{Fixing the deformed Kerr solution}

As we have noted, the deformation function $W(r,\th)$ is fixed by the requirement of the stress-tensor to be that of an anisotropic relativistic fluid. It is achieved with solving for the following equation on the deformation function:
\be
\fr{r-M}{\D(r)} \pa_\th W(r,\th)+\cot\th \,\pa_r W(r,\th)=0.
\la{G23=0}
\ee
One of the solutions to eq. \rf{G23=0} is
\be
W_1(r,\th)=C_1 \ln\left(r_0\D^{-1/2}\,\sin\th \right),
\la{Sol1}
\ee
with some constant parameter $r_0$ of the length dimensionality.
By use of separation of variables, another solution to \rf{G23=0} comes as follows:
\be
W_2(r,\th)=C_2\, \D^{-l/2} \sin^l\th.
\la{Sol2}
\ee
Here $C_{1,2}$ and $l$ are the integration and separation constants, respectively. The general solution to eq. \rf{G23=0} becomes
\be
W(r,\th)=W_1(r,\th)+W_2(r,\th)=C_1 \ln\left(r_0 \D^{-1/2}\,\sin\th \right)+C_2\, \D^{-l/2} \sin^l\th .
\la{Solgen}
\ee

Now we have to specify three of four unknown parameters in the solution \rf{Solgen}: $C_{1,2}$ and $l$. As a criterion, we will use the assumption of the same or nearly the same types of singularities for the generalized metric as for the standard Kerr metric. Hence, we have to compute the Kretschmann scalar for the metric \rf{Rotellipsm} with the deformation function \rf{Solgen} to explore singularities of our model in dependence on different choices of the integration and separation constants.

Let's recall the singularities of the standard Kerr metric first. For the Kerr solution the Kretschmann scalar becomes \cite{Henry:1999rm}
\be
K_{\text{ Kerr}}=\fr{48 M^2}{\r^{12}} \left(r^6 - 15 a^2 r^4 \cos^2\th + 15 a^4 r^2 \cos^4\th - a^6 \cos^6\th \right).
\la{KKerr}
\ee 
According to \rf{KKerr}, the true singularity of the Kerr metric is located at $\{r=0,\th=\pi/2\}$. Also, the Kretschmann scalar is an even function under the $\th$ inversion $K(\th)=K(-\th)$. And it is regular in the north and south pole directions $\th=0$ and $\th=\pi$. We would like to keep these properties for the generalized Kerr metric \rf{Rotellipsm} as much as possible, that would properly constraint the unfixed parameters of $W(r,\th)$ function.

The Kretschmann scalar for the line element \rf{Rotellipsm} turns out to be 
\be
K=e^{-4 W(r,\th)}K_{\text{Kerr}}+F(W(r,\th), r,\th;M,a),
\la{KKerrgen}
\ee
with an additional function $F(W(r,\th), r,\th;M,a)$, the exact form of which is given in Appendix \ref{app:F-exact}. The structure of \rf{KKerrgen} is such that the trivialization $W(r,\th)$ corresponding to the trivial (zero) integration constants returns us, as expected, to the standard Kretschmann scalar \rf{KKerr}. Hence, our task is to study different non-trivial choices of $C_{1,2}$.

Let's begin with $C_2=0$. In this case 
\be
W(r,\th)=C_1 \ln\left(r_0\D^{-1/2}\,\sin\th \right),\qquad \exp\left[ 2C_1 \ln\left(r_0\D^{-1/2}\,\sin\th \right)\right]=\fr{(r_0\sin\th)^{2C_1}}{\D^{C_1}}.
\la{f1}
\ee
Then, we get the following expression for the line element:
\be
\begin{split}
ds^2=-\fr{\D}{\r^2}\left(dt-a\sin^2 \th \,d\vf\right)^2&+\fr{\r^2(r_0\sin\th)^{2C_1}}{\D^{1+C_1}} dr^2+\fr{\r^2(r_0\sin\th)^{2C_1}}{\D^{C_1}} d\th^2\\
&+\fr{\sin^2 \th}{\r^2}\left[\left(r^2+a^2\right)d\vf-a dt \right]^2.
\end{split}
\la{genKerrC1}
\ee
Naively, it seems that with $C_1=-1$ we could avoid the singularity on the horizon, though it would appear new singular points at $\th=\{0,\pi\}$. In fact, at such choice of the deformation function, the Kretschmann scalar has the form of the product of the singular at some values of $\{r,\th\}$ pre-factor on a smooth, non-factorizable in $\D$ and $\r$, function of these variables:
\be
K_{\big|_{C_1=-1}}=\fr{4r^4_0}{\D^4 \r^{12}}\,{\cal F}_{(-1)}(r,\th;M,a).
\la{K-1}
\ee
Therefore, in addition to the standard points of the true singularity for the Kerr metric, we have the physical singularity on the horizon, where $\D=0$. It is interesting to note that, according to \rf{K-1}, there is no singularity at $\th=\{0,\pi\}$, though, at first sight, the metric \rf{genKerrC1} indicates it. This is a good example of how one should not draw conclusions about the nature of singularities based only on the exact form of the metric.

For $C_1=+1$ we have another interesting feature of the metric \rf{genKerrC1}. 
In this case the Kretschmann scalar turns out to be
\be
K_{\big|_{C_1=+1}}=\fr{4}{r^4_0\r^{12}\sin^8\th} {\cal F}_{(+1)}(r,\th;M,a),
\la{K+1}
\ee
where, as in the previously considered case with $C_1=-1$, ${\cal F}_{(+1)}(r,\th;M,a)$ is another non-factorizable in $\D$ and $\r$ smooth function of coordinates and black hole parameters. Here we are faced with the opposite situation: according to the structure of the Kretschmann scalar \rf{K+1}, the metric
\be
\begin{split}
ds^2=-\fr{\D}{\r^2}\left(dt-a\sin^2 \th \,d\vf\right)^2&+\fr{\r^2 r^2_0\sin^{2}\th}{\D^{2}} dr^2+\fr{\r^2 r^2_0\sin^{2}\th}{\D} d\th^2\\
&+\fr{\sin^2 \th}{\r^2}\left[\left(r^2+a^2\right)d\vf-a dt \right]^2
\end{split}
\la{genKerrC1+1}
\ee
has not physical singularity on the horizon. However, it possesses the additional, in compare to Kerr, true singularity in the north and south pole directions. In astrophysics, the equatorial plane with $\th=\pi/2$ is usually reserved for the direction of the accretion disk of external matter, while the north and south pole directions $\th=\{0,\pi\}$ are associated with jets of ultra high-energy particles produced by a black hole.

Next, let's study another part of the solution \rf{Solgen} corresponding to the case of $C_1=0$. Here we get
\be
W(r,\th)=C_2\, \D^{-l/2}\sin^l\th.
\la{f2}
\ee
For dimensionality reasons, the dimensionality of $C_2$ shall compensate the dimensionality of $\D^{-l/2}$. To make a proper choice of the separation constant, consistent with the symmetry properties of the standard Kerr solution, we have to examine the Kretschmann scalar for a general value of $l$. The explicit expression for this quantity is given in Appendix \ref{app:Kretsch-details}.

A brief inspection of the obtained for the Kretschmann's scalar expression \rf{KC2gen} shows that the symmetry with respect to $\th \ra -\th$ in the standard Kerr solution, which we would like to preserve, can be achieved for $l=2n$, $n \in \mathbb{Z}$. Then, computing the Kretschmann scalar for positive and even $l$s, one may notice that this quantity is structured as follows, 
\be
K_{l>0}=\fr{4\exp(-4\tilde{C}_2 r^l_0\sin^l\th/\D^{l/2})}{\D^{l+2}\r^{12}}\, {\cal F}_{(l>0)}(r,\th;M,a),
\la{Kl>0gen}
\ee
with some regular and non-factorizable in $\D$ and $\r$ function ${\cal F}_{(l>0)}(r,\th;M,a)$ and a dimensionless integration constant $\tilde{C}_2$. As a result, the Kretschmann scalar shows physical singularity on the horizon just in the north and south pole directions. (For $\th=\{0,\pi\}$ we have a pure singularity on the horizon; for other values of $\th$ we get strong exponential suppression for positive $\tilde{C}_2$ indicating the absence of gravity in all but $\th=\{0,\pi\}$ directions that looks absolutely unphysical.)

For negative admissible values of $l$, the Kretschmann scalar possesses another interesting feature. For instance, for the simplest choice of $C_2=1/r_0$ and $l=-2$, the Kretschmann scalar becomes: 
\be
K_{l=-2}=\frac{16 e^{-\frac{4 \D }{r_0^2 \sin ^2 \theta}}}{r_0^4\r^{12}\sin^8\th}
\, {\cal F}_{(l=-2)}(r,\th;M,a).
\la{Kl-2}
\ee
As one can see, the true singularity of the solution is located at $r=0$ and $\th=\pi/2$ as in the standard Kerr solution, and the additional points of singularity, located in the north and south pole directions, are exponentially suppresed.

For a general function $W(r,\th)$ from \rf{Solgen}, corresponding to the standard singularities of the Kerr metric with exponentially suppressed the north/south Pole singularities, the generalized metric \rf{Rotellipsm} looks as follows ($C_2=\tilde{C}_2 r^l_0$):
\be
\begin{split}
ds^2=&-\fr{\D}{\r^2}\left(dt-a\sin^2 \th \,d\vf\right)^2+\fr{\exp\left({\fr{2\tilde{C}_2 r^l_0\sin^l \th}{\D^{l/2}}}\right)\r^2(r_0\sin\th)^{2C_1}}{\D^{1+C_1}} dr^2\\
&+\fr{\exp\left({\fr{2\tilde{C}_2 r^l_0\sin^l \th}{\D^{l/2}}}\right)\r^2(r_0\sin\th)^{2C_1}}{\D^{C_1}} d\th^2+\fr{\sin^2 \th}{\r^2}\left[\left(r^2+a^2\right)d\vf-a dt \right]^2;\\
&C_1\ge 0,\qquad \tilde{C}_2 \ge 0,\qquad C_1+\tilde{C}_2 \ne 0, \qquad l=-2n,\,\,n\in \mathbb{N}.
\end{split}
\la{genKerrC1C2}
\ee

Let's summarize our findings. We have found a solution for a rotating black hole immersed in external matter. The character of this solution is such that it preserves the symmetry and singularity properties of the standard Kerr, with one exception. Namely, there are other points of singularity, located at the north and south poles of the black hole. Moreover, it should be noted that, depending on the choice of the deformation function, this additional singularity of the metric can be pronounced (as for a solution of type \rf{f1}) or smoothed (exponentially suppressed as for a solution of type \rf{f2}). For the general type metric \rf{genKerrC1C2} the additional singularity is smoothed. The exponential suppression at the points of additional singularity results in a near-zero value of the Kretschmann scalar there. It means that gravity is weak at the north and south poles, and particles from the black hole ergoregion may freely escape the black hole in the north and south pole directions, forming a well-known picture of outgoing jets. However, we must be sure that the external matter included in the construction is physical, and does not suffer from different pathologies.

\subsection{NEC, WEC, SEC, DEC and FEC}

Now we want to check the standard requirements on the stress-energy tensor of matter for the established solution (see a comprehensive discussion of energy conditions in Ref. \cite{Curiel:2014zba}). 

The null energy condition (NEC) is
\be
T_{mn}l^m l^n \ge 0,
\la{NEC}
\ee
where $l^m$ is one of two null-vectors for the geometry with metric \rf{Rotellipsm}. Following Ref. \cite{Gourgoulhon:2005ng}, we can construct this null-vector as $l^m=1/N\times(u^m+n^m)$, where $N$ is a non-trivial normalization factor, $u^m$ has been defined in eqs. \rf{udef}, and the vector $n^m$ is as follows \cite{Nurmagambetov:2022wcw}:
\be
n^m=\left(0,\fr{\tilde{\r}^2}{\D},0,0\right).
\la{ndef}
\ee

For the general form \rf{Solgen} of the deformation function $W(r,\th)$ we get the following result:
\[
T_{mn} l^m l^n=-\fr{1}{8\pi G \,N^2 \sin^2\th\, \r^2}\exp\left(-2C_2 \fr{\sin^l\th}{\D^{l/2}}\right)\fr{\D^{C_1-l/2-1}}{\sin^{2C_1}\th} \times
\]
\be
\times \left[a^2 \left(-2C_1 \sin^2\th \D^{l/2}+C_2 l \sin^l\th \left((l+2)\cos^2\th-2 \right)\right) \right.
\la{NEC1Solgen}
\ee
\[
\left. +2C_1 M^2 \sin^2\th \D^{l/2}+C_2 l \sin^l \th \left((l+2)M^2 \sin^2\th-2lMr+lr^2 \right) \right].
\]
Fixing the parameters as (cf. eq. \rf{genKerrC1C2}) $C_1=1$, $C_2=1/r_0$, $l=-2$, we get\footnote{Any other admissible choice of the parameters does not change the qualitative behavior of the NEC.}
\be
T_{mn} l^m l^n=-\fr{2}{8\pi G \,N^2 r^2_0 \r^2 \sin^6\th}\exp\left(-\fr{2\D}{r^2_0 \sin^2\th} \right)\left[(M^2-a^2)r^2_0 \sin^4 \th+2\D^2 \right].
\la{NEC2}
\ee
Therefore, we conclude that the NEC is violated in the vicinity of and far from the black hole, turning to zero at the radial infinity. However, the NEC is not a reliable condition in verifying the physicality of matter fields.

Next, we verify the weak energy condition (WEC). The condition reads
\be
{\cal E}=T_{mn} u^m u^n \ge 0 .
\la{WEC}
\ee
With taking into account the expression for a time-like vector field $u^m$ (cf. eq. \rf{udef}) and the line element \rf{genKerrC1C2}, we have
\be
\begin{split}
&{\cal E}=-\fr{\D^{C_1-1-l/2}}{8\pi G\r^2\sin^{C_1+2}\th}\exp\left(-\fr{2C_2 \sin^l\th}{\D^{l/2}}\right)\times\\
\times
&\Big[a^2\left(C_1(\cos^2\th-2)\D^{l/2}+C_2 l \sin^l\th\left((l+1)\cos^2\th-2 \right)\right)\\
&-C_1\left(\D-a^2-M^2\sin^2\th\right)\D^{l/2}+C_2 l \sin^l \th \left((l-1)(\D-a^2)+(l+1)M^2\sin^2\th \right) \Big].
\end{split}
\la{rhoC1C2}
\ee
Fixing the parameters as before ($C_1=1$, $C_2=1/r_0$, $l=-2$), we arrive at the following expression for the energy density:
\be
\begin{split}
{\cal E}=-\fr{1}{8\pi G r^2_0 \r^2 \sin^6\th}\exp\left(-\fr{2\D}{r^2_0 \sin^2\th} \right)&\left[ \left((M^2-a^2)\sin^2\th-\D \right)r^2_0+4\D^2 \right.\\
&\left.+2\D\left(\D+(M^2-a^2)\sin^2\th \right) \right].
\end{split}
\la{WECSol}
\ee
Since the WEC is important from the physical point of view, and, moreover, it is expected that matter near a Super-Massive Black Hole is of the positive energy density (see, e.g., Ref. \cite{Viaggiu:2023ewh} in this respect), we have to look at this condition in more detail.

First, let's consider the case of extremal (a fast rotating) black hole, when $M=\pm a$. Here we can use the still unfixed parameter of the solution $r_0$ to make the r.h.s. of \rf{WECSol} non-negative. It is achieved by $r^2_0 \ge 6\D$. For a non-extremal black hole the energy density becomes non-negative for 
\be
r^2_0 \ge (6\D^2+2\D(M^2-a^2)\sin^2\th)/(\D-(M^2-a^2)\sin^2\th),
\la{r0WEC}
\ee
that sets another constraint on the location of matter fields in the radial direction:
\be
\D>(M^2-a^2) \qquad \leadsto \qquad r_{\pm}>M\pm \sqrt{2(M^2-a^2)}.
\la{rpmWEC}
\ee
The smaller root of $r_\pm$ is chosen for $1/2<(a/M)^2\le 1$. The largest root $r_+$ works in the case when $0\le (a/M)^2 \le 1/2$. In both cases $r_{\pm}<r^{(-)}_{\text{MBCO}}$, where $r^{(-)}_{\text{MBCO}}$ is the smaller radius of the marginally bound circular orbit (MBCO) \cite{Bardeen:1972fi} (see also Ref.\cite{Kopacek:2024pfd})
\be
r_{\text{MBCO}}=2M\mp a+2\sqrt{M\pm a} .
\la{MBCO}
\ee
Since any physically relevant matter can form bound circular orbits only above the values of $r_{\text{MBCO}}$, the energy density \rf{WECSol} when \rf{r0WEC} is chosen will be definitely positive under such conditions. 

Concerning the strong energy condition (SEC),
\be
\left(T_{mn}-\fr12 g_{mn}T \right)u^m u^n \ge 0,
\la{SEC}
\ee
it is satisfied everywhere in the outer, w.r.t. the black hole horizon, space, since the l.h.s. of \rf{SEC} is equal to zero.

The other popular energy condition is the dominant energy condition (DEC) (see Ref.  \cite{Curiel:2014zba}). It states that the vector $T_{mn} u^n$ has to be a time-like vector. Since this condition is weak in compare to the abovementioned ones (see the discussion in Ref. \cite{Abreu:2011fr} in this respect), it is better to use either NEC, WEC and SEC, or alternate, more physically relevant, energy conditions.

Now let's check the flux energy condition (FEC), which was proposed in  \cite{Abreu:2011fr} in the cosmological context and ruled out the possibility of tachyonic matter. This condition reads $T_{mn} \xi^n$ is causal (that is, time-like or null) for any time-like vector $\xi^m$ (which, in our case, can be chosen to be $u^m$), or, equivalently,
\be
T_{mn} {T^m}_k u^n u^k \le 0.
\la{FEC}
\ee
For the line element \rf{genKerrC1C2} with $C_1=1$, $C_2=1/r_0$, $l=-2$, the FEC becomes
\be
\begin{split}
&T_{mn} {T^m}_k u^n u^k =-\fr{1}{(8\pi G)^2 r^4_0 \r^4 \sin^{12}\th}\exp\left(-\fr{4\D}{r^2_0 \sin^2\th} \right)\times\\
&\times\left[\left((M^2-a^2)\sin^2\th-\D \right)r^2_0+4\D^2+2\D\left(\D+(M^2-a^2)\sin^2\th \right) \right]^2.
\end{split}
\la{FEC1}
\ee
Apparently, this condition is satisfied everywhere in the outer, w.r.t. the black hole horizon, space.

\ssk
Thus, we have constructed the solution to Einstein equations for a rotating black hole in the presence of matter fields. It has been found conditions and constraints on the solution parameters leading to a physically viable stress-energy tensor satisfying most popular energy conditions. The next natural step in our research program is to generalize the found solution to the case of constant curvature spaces, which play an important role in Cosmology and Holography.

\subsection{(A)dS generalization of the deformed Kerr}

Recall that the (A)dS generalization of the standard Kerr solution \cite{Carter:1972,Hawking:1998kw} is 
\be
\begin{split}
ds^2=&-\fr{\D_r}{\r^2}\left(dt-\fr{a \sin^2\th d\vf}{\Xi} \right)^2+\fr{\r^2}{\D_r} dr^2+\fr{\r^2}{\D_\th} d\th^2\\
&+\fr{\D_\th \sin^2\th}{\r^2}\left(a dt-\fr{(r^2+a^2)d\vf}{\Xi} \right)^2 .
\end{split}
\la{KerrAdS}
\ee
Here
\be
\begin{split}
\D_r=\left(1+\fr{r^2}{L^2} \right)&(r^2+a^2)-2Mr,\qquad \D_\th=1-\fr{a^2}{L^2}\cos^2\th \\
&\r^2=r^2+a^2\cos^2\th,\qquad \Xi=1-\fr{a^2}{L^2},
\end{split}
\la{Ddefs}
\ee
and the characteristic length of four-dimensional (A)dS space is related to the cosmological constant:
\be
L^2 \equiv -\fr{3}{\L}\,\,\,\,(\text{AdS}_4),\qquad L^2 \equiv \fr{3}{\L}\,\,\,\,(\text{dS}_4).
\la{Ldef}
\ee

On account of our previous experience in constructing the deformed Kerr solution \rf{Rotellipsm}, the proposed extension to (A)dS space is of the form
\be
\begin{split}
ds^2=&-\fr{\D_r}{\r^2}\left(dt-\fr{a \sin^2\th d\vf}{\Xi} \right)^2+\fr{\tilde{\r}^2}{\D_r} dr^2+\fr{\tilde{\r}^2}{\D_\th} d\th^2\\
&+\fr{\D_\th \sin^2\th}{\r^2}\left(a dt-\fr{(r^2+a^2)d\vf}{\Xi} \right)^2,\qquad \tilde{\r}^2=e^{2W(r,\th)}\r^2.
\end{split}
\la{KerrAdSm}
\ee

Straightforward computations show that the line element \rf{KerrAdSm} solves the Einstein equations with a cosmological constant
\be
G_{mn}+g_{mn}\L=8\pi G T_{mn},
\la{EinsL}
\ee
and non-trivial components of the stress-energy tensor of matter are presented in Appendix \ref{app:deformed-Kerr-SET}.

In order to reduce the stress-energy tensor of matter to that of an anisotropic relativistic fluid, it is necessary to fix the deformation function as
\be
\left(L^2(r-M)+r(2r^2+a^2)\right)\D_\th \,\pa_\th W+\D_r \left(\D_\th L^2+a^2\sin^2\th \right)\cot \th \,\pa_r W=0.
\la{G23=0dS}
\ee
Apparently, in the limit of small cosmological constant $\L \ra 0$ (or, equivalently, $L \ra \infty$), eq. \rf{G23=0dS} turns into eq. \rf{G23=0}.

Solving for eq. \rf{G23=0dS}, we get
\be
W(r,\th)=C_1 \arccosh\left[-\fr1{\D_r\D_\th \sin^2\th}\left(\fr{\D_r^2}{4r^2_0}+r^2_0-r^2_0\cos^2\th \left(\D_\th+\fr{a^2}{L^2} \right)\left(1+\D_\th \sin^2\th \right)\right) \right],
\la{Sol1AdS}
\ee
where the parameter $r_0$ has been introduced as before to make the expression inside the square brackets dimensionless.

The other solution can be obtained by use of separation of variables, that leads to
\be
W(r,\th)=C_2 \left(\fr{\sin^2\th \D_\th}{\D_r} \right)^{\fr1{2} L^2 l}
\la{Sol2AdS}
\ee
Hence, the analog of the solution \rf{Solgen} for (A)dS space turns out to be
\[
W(r,\th)=C_1 \arccosh\left[-\fr1{\D_r\D_\th \sin^2\th}\left(\fr{\D_r^2}{4r^2_0}+r^2_0-r^2_0\cos^2\th \left(\D_\th+\fr{a^2}{L^2} \right)\left(1+\D_\th \sin^2\th \right)\right) \right ]
\]
\be
+C_2 \left(\fr{\sin^2\th \D_\th}{\D_r} \right)^{\fr1{2} L^2 l} .
\la{SolgenAdS}
\ee

If the correspondence between equations \rf{G23=0dS} and \rf{G23=0} for the deformation function $W(r,\th)$ in (A)dS and flat spacetimes is direct, we can not say the same for the solutions \rf{SolgenAdS} and \rf{Solgen}. We refer the reader to Appendix \ref{app:Zero-lambda-W}, where we collect technical details on the zero cosmological constant limit of the solution \rf{SolgenAdS}. 

Concerning the choice of the unconstrained constants $C_{1,2}$ and $l$ in the solution \rf{SolgenAdS}, the first two are fixed from the requirement of smooth zero cosmological constant limit.\footnote{Roughtly, the deformation function for a small cosmological constant almost coincides with the deformation function in the flat space.} The additional requirement on the separation constant $l$ to be negative and even naturally follows from a more accurate analysis of this limit (see Appendix \ref{app:Zero-lambda-W} for details).

Having established viable solutions for stationary rotating black holes in the matter fields environment, let us proceed with extending the construction to the non-stationary case of Kerr-Vaidya black holes. But first we have to construct an analog of Vaidya's black hole on this route, hence, we have to turn to the non-rotating limit of the deformed Kerr solution.

\section{The non-rotating limit of the deformed Kerr black hole}

The non-rotating limit of a deformed Kerr black hole is important for at least two reasons. First, in our ``top-bottom'' approach, the corresponding Schwarzschild-type solution has not yet been analyzed. And second, it is not so clear, how properties of the solution depend on the coordinate choice, being crucial for further construction of the non-stationary version of the deformed Kerr.

\subsection{Schwarzschild-type solution with external matter}

We begin with a Schwarzschild-type solution in usual (Boyer-Lindquist from the Kerr solution point of view) coordinates. In the non-rotating limit $a \ra 0$ the line element \rf{Rotellipsm} turns into
\be
ds^2=-f(r)dt^2+\fr{e^{2W(r,\th)}}{f(r)}+r^2 e^{2W(r,\th)}d\th^2+r^2\sin^2\th d\vf^2,\qquad f(r)=1-\fr{2M}{r}.
\la{SSdef}
\ee
Apparently, spherical symmetry is broken here, and the Einstein equations for the metric \rf{SSdef} signal a nontrivial stress-energy tensor of matter. To write down the r.h.s. of the Einstein equations as the stress-energy tensor of an anisotropic fluid, one has to require the $G_{r\th}$ component of the Einstein tensor to vanish. Hence, 
\be
G_{r\th}=\left(\fr1{r}+\fr{\pa_r f}{2f} \right) \pa_\th W+\cot \th \pa_r W=0.
\la{G23SS}
\ee
Solving for eq. \rf{G23SS}, we get 
\be
W(r,\th)=C_1 \ln \left(\fr{r_0\sin\th}{rf^{1/2}}\right),
\la{C1SolSS}
\ee
which is the $a \ra 0$ limit of the solution \rf{f1}. (For the sake of simplicity, we will focus only on this kind of solutions to eq. \rf{G23SS}.)

With taking into account \rf{C1SolSS}, the elements of the stress-energy tensor \rf{Tmnan} comes as follows:
\be
{\cal E}=\fr{C_1 e^{-2W}}{8\pi G r^4}\left(\fr{r^2 f}{\sin^2 \th}-M^2 \right),
\la{ESS}
\ee
\be
P=-\fr{C_1 e^{-2W}}{8\pi G r^4 f}\left(\left(M-r \right)^2+r^2 f \cot^2\th \right),
\la{PSS}
\ee
\be
P_\th=-P,\qquad P_\vf=-{\cal E}.
\la{PthvfSS}
\ee
Clearly, the stress-energy tensor becomes traceless in the case.

It is instructive to take a look on various energy conditions for the deformed Schwarzschild-type solution \rf{SSdef}, \rf{C1SolSS}. By use of the ``top-bottom'' correspondence between solutions for rotating and non-rotating black holes, we fix $C_1>0$, then the energy \rf{ESS} is positive (the WEC) if
\be
r^2 \left(1-\fr{2M}{r} \right)-M^2 \ge 0 \quad \rightleftarrows \quad r\ge (1+\sqrt{2})M.
\la{Epos}
\ee 
The equatorial circular orbits in this kind of solution are possible for $r\ge 3M$, where equality corresponds to photons.\footnote{In a very rough picture of an accretion disk around a black hole, the circular orbits of photons determine the nearest location of matter to the event horizon. The innermost stable cirular orbits (ISCO) for equatorial time-like geodesics are located at $r=6M$, see Ref. \cite{Song:2021ziq}.} Hence, the radial locations of particle orbits corresponding to the positivity of their energies fall into the physically admissible range. Note also the curious coincidence of the bound value $r=(1+\sqrt{2})M$ with the radius of polar circular photon orbits in the case of extremal Kerr black hole (see, e.g., \cite{Kopacek:2024pfd} in this respect).

The NEC, see \rf{NEC}, is always realized in this case for a positive $C_1$:
\be
8\pi G \,T_{mn} l^m l^n=\fr{2C_1 M^2}{r^4 f}\left(\fr{\sqrt{f}}{r_0 \sin\th}\right)^{2C_1} \ge 0.
\la{NECSS}
\ee
Here we have used the null-vector $l_m=(\sqrt{f},e^{W}/\sqrt{f},0,0)$. The SEC, due to the stress-energy tensor traceless condition, coincides with the NEC. Finally, the FEC looks as
\be
-{\cal E}^2 \le 0,
\la{FECSS}
\ee
that can always be realized. Therefore, we can conclude that for physically relevant configurations, corresponding to circular orbits of matter particles with the radial locations $r\ge 3M$, the solution describes a static black hole with ordinary (non-tachyonic) matter.

\subsection{A generalized static black hole in the null coordinates}

Now, let's consider properties of the solution \rf{SSdef} in the null (Finkelstein-Eddington) coordinates. To involve the null coordinates in the game it is not simply to rewrite the established characteristics and parameters of the solution, since it is not clear what changes the stress-energy tensor of matter will undergo at such replacement of coordinates. However, the benefits of such transform are obvious, since it opens a way to construct a generalization of the Vaidya solution for a non-stationary black hole.

So, let's re-organize the line element \rf{SSdef} in terms of $(v,r,\th,\vf)$, where three of four coordinates are the same as in \rf{Rotellipsm} (in the $a\ra 0$ limit), while the ingoing coordinate $v=t+r^*$ includes the ``tortoise'' coordinate $r^*$, related to $r$ via $dr^*=dr/(1-2M/r)$. Then, by use of standard manipulations, one gets
\be
ds^2=-f(r)dv^2+2 e^{W(r,\th)}dv dr+r^2 e^{2W(r,\th)} d\th^2+r^2 \sin^2\th d\vf^2,\quad f(r)=1-\fr{2M}{r},
\la{SSm}
\ee    
which corresponds to a generalization of the Schwarzschild solution in the presence of matter. From the Einstein equations \rf{Eins} it follows that the stress-energy tensor of matter has the complete structure of a viscous fluid stress-energy tensor \rf{Tmndef} with
\be
{\cal E}=\fr{e^{-2W}}{32\pi G r^2} \left[4+3\left(\pa_\th W\right)^2-4 \pa^2_\th W-2r \pa_r f(2+r \pa_r W)-4f(1+r\pa_r W+r^2 \pa^2_r W) \right];
\la{ESSEF}
\ee
\be
\begin{split}
P=\fr{e^{-2W}}{96 \pi G r^2} \left[-4+4r^2 \pa^2_r f+(\pa_\th W)^2+4 \pa^2_\th W
\right.\\
\left.+2r \pa_r f(6+r \pa_r W)+4 f(1+r\pa_r W+r^2\pa^2_r W) \right];
\end{split}
\la{PSSEF}
\ee
\be
\begin{split}
&q_m=\left(0,q_r,q_\th,0 \right),\\
q_r=-\fr{e^{-W}}{16\pi G r^2 f^{1/2}}&\left(\cot \th \pa_\th W-(\pa_\th W)^2+\pa^2_\th W \right),\\
q_\th=\fr{e^{-W}}{16\pi G f^{1/2}}&\left(\pa_r f\pa_\th W+f\left(-\pa_\th W \pa_r W+\pa_\th \pa_r W \right)\right);
\end{split}
\la{qmSSEF}
\ee
\[
\t_{mn}=\fr{1}{8\pi G}\left(
\begin{array}{cccc}
0&0&0&0\\
0& \tilde{\t}_{rr} & \tilde{\t}_{r\th} & 0\\
0& \tilde{\t}_{\th r} & \tilde{\t}_{\th\th}&0\\
0&0&0& \tilde{\t}_{\vf\vf}
\end{array}
\right),
\]
\be
\begin{split}
\tilde{\t}_{rr}=\fr1{4r^2f}\left[-4-4\cot\th \pa_\th W+(\pa_\th W)^2+4f(1+r \pa_r W)+2 r \pa_r f(2+r\pa_r W) \right],\\
\tilde{\t}_{r\th}=\left(\fr1{r}+\fr{\pa_r f}{2f}\right)\pa_\th W+\cot\th \pa_r W,
\\
\tilde{\t}_{\th\th}=\fr{1}{4} \left[2r^2\pa^2_r f+4\cot\th\pa_\th W+(\pa_\th W)^2-4r f\pa_r W-2r\pa_r f(-2+r\pa_r W) \right],\\
\tilde{\t}_{\vf\vf}=\fr14 e^{-2W}\sin^2\th \left[2r^2 \pa^2_r f-(\pa_\th W)^2+4\pa^2_\th W+4r f \pa_r W \right.
\\
\left. +2r\pa_r f (2+r\pa_r W+4r^2 f\pa^2_r W) \right].
\end{split}
\la{tauSSEF}
\ee
Such a complete form of the stress-energy tensor is associated with certain difficulties in fixing the deformation function $W(r,\th)$ as an analytic function of two variables directly, as a solution to the corresponding differential equations.

Indeed, to reach the structure of the stress-tensor of an anisotropic relativistic fluid in the case of metric \rf{SSm}, one needs to set to zero the following components of the Einstein tensor: 
\be
G_{v\th}=\fr{e^{-W}}{2r^2}\left[\pa_\th W(2M-(r-2M)r\pa_r W)+r(r-2M)\pa_\th \pa_r W \right],
\la{G13SSEF}
\ee
\be
G_{r\th}=\fr1{2r} \left[\pa_\th W(2+r\pa_r W)+r(2\cot\th \pa_r W-\pa_\th \pa_r W) \right].
\la{G23SSEF}
\ee
Our attempts to find the desired analytical solution for both $G_{v\th}=0$ and $G_{r\th}=0$ have failed.\footnote{For instance, $G_{v\th}=0$ results in the deformation function depending solely on the radial coordinate, that is unacceptable within the developed here approach. Solving for $G_{r\th}=0$, one can neither find an analytic solution to the equation, nor use the separation of variables method.} Hence, in this case we keep the complete structure of the stress-tensor of matter, and have to fix the deformation function from other considerations.   

We can use the following line of reasoning to this end. The considered here spacetime configuration is related to the geometry of \rf{SSdef} via a smooth coordinate transformation. And the ``old'' coordinates $\{r,\th\}$ are not affected by this transformation; they coincide with ``new'' $\{r,\th\}$ coordinates. It means that the form of the previously established deformation function, eq. \rf{C1SolSS}, does not change. And we can apply it ``by hands'' to finally fix the metric \rf{SSm} and to investigate different energy conditions for matter fields.

Let's begin with the NEC. The corresponding null-vector in the current spacetime geometry is
\be
l^m=\left(e^{-W},\fr12 fe^{-2W},0,0\right),
\qquad l^m l_m =0.
\la{nullSSEF}
\ee
Hence, the NEC, for $C_1=1$, becomes as follows:
\be
T_{mn}l^m l^n\ge 0 \quad \rightleftarrows \quad \left(3r^2-6Mr-2M^2+\left(r^2-2Mr+2M^2 \right)\cos 2\th \right) \ge0.
\la{NECSSEF}
\ee
And this condition always satisfies when $r\ge 2M$. With the same rationales as in the case of inequality \rf{Epos}, we can conclude that physical matter fields form the configuration corresponding to the NEC fulfillment. 

For the WEC, again with $C_1=1$, we get the same result as in \rf{Epos}. (That justifies this choice of the integration constant.) The SEC \rf{SEC} becomes as
\be
\left(1-\fr{2M}{r} \right)\fr{\cos^2\th}{\sin^4 \th} \ge 0,
\la{SECSSEF}
\ee  
which is realizable for any value of $\th$. Note that for $\th=\pi/2$, corresponding to the accretion disk direction, the energy density becomes equal to zero. Formally, this means that the particles are ``frozen'' with respect to the time coordinate $v$, but it is impossible to ``freeze'' photons. Hence, in this model the matter content of the accretion disk consists of massive particles.  

What concerns the DEC and the FEC, these energy conditions turn out to be the most difficult to analyze. These conditions are summarized as
\be
\begin{split}
\text{FEC}=\fr{1}{\left(32\pi G r^2 r^2_0\sin^2\th \right)^2}\Bigg[-16M^4+56M^2 r-48 M^2 r^2+36 M r^3-13 r^4\\
+\fr{r(r-2M)}{\sin^2\th} \left(60M^2-76M r+46 r^2-\fr{45r(r-2M)}{\sin^2\th} \right)\Bigg]\le 0\,.
\end{split}
\la{FECEF}
\ee
When $\sin^2\th=1$, the inequality is definitely realized for $r\ge 3M$, where equatorial circular orbits of any type of physical matter (massive or massless) are located.

\subsection{Vaidya-type solution}

As a step toward a Kerr-Vaidya-type solution for a non-stationary rotating black hole in the presence of matter, let's consider the extension of metric \rf{SSm} to that of a Vaidya-type \cite{Vaidya:1951zz,Vaidya:1951zza} black hole. In this case we focus on the following line element\footnote{In the most general case we have to consider the deformation function depending on time-like coordinate as well. Additional reasons for including this dependency will become clear further on.}
\be
\begin{split}
ds^2=-{\cal F}(v,r)dv^2+&2 e^{W(v,r,\th)}dv dr+r^2 e^{2W(v,r,\th)} d\th^2+r^2 \sin^2\th d\vf^2, \\
&{\cal F}(v,r)=1-\fr{2M(v)}{r}.
\end{split}
\la{Vm}
\ee
For the same reasons as in case of the seed metric \rf{SSm}, the stress-energy tensor of matter keeps the full structure of  a viscous fluid stress-energy tensor. To compute different physical quantities of \rf{Tmndef}, we have to use the 4-velocity vector
\be
u^m=\left(-\fr{1}{\sqrt{1-\fr{2M(v)}{r}}},0,0,0\right),
\la{uVmdef}
\ee
and the projector onto a 3-dimensional manifold $h_{mn}=g_{mn}+u_m u_n$. Then we have
\be
\begin{split}
{\cal E}=&\fr{e^{-2f}}{32\pi G r^2} \Bigg[\fr{1}{\cal F}\left(-2 e^{W} r \pa_v {\cal F} \left(2+r \pa_r W \right) \right)+4+3\left(\pa_\th W\right)^2 \\
&
-4 \pa^2_\th W-2r \pa_r {\cal F}(2+r \pa_r W)-4{\cal F}(1+r\pa_r W+r^2 \pa^2_r W) \Bigg]\\
&+\fr{e^{-W}}{16\pi G r^2{\cal F}}\left(r^2 \pa_r {\cal F}-2 r {\cal F}(1+r\pa_r W) \right)\pa_v W-\fr1{4\pi G} e^{-W}\pa_v \pa_r W-\fr1{8\pi G {\cal F}}\pa^2_v W;
\end{split}
\la{EVm}
\ee
\be
\begin{split}
P=\fr{e^{-2W}}{96 \pi G r^2} \Bigg[\fr1{{\cal F}}\left( -2e^{W}  r \pa_v {\cal F} \left(2+r \pa_r W\right) \right)-4+4r^2 \pa^2_r {\cal F}+(\pa_\th W)^2+4 \pa^2_\th W\\
+2r \pa_r{\cal F}(6+r \pa_r W)+4{\cal F}(1+r\pa_r W+r^2\pa^2_r W) \Bigg]\\
+\fr{e^{-W}}{96\pi G r^2{\cal F}}\left(20 r{\cal F}+2 r^2 \pa_r {\cal F}+12 r^2 {\cal F}\pa_r W \right)\pa_v W+\fr1{12\pi G} e^{-W}\pa_v \pa_r W-\fr1{24\pi G {\cal F}}\pa^2_v W;
\end{split}
\la{PVm}
\ee
\be
\begin{split}
&q_m=\left(0,q_r,q_\th,0 \right),\\
q_r=&-\fr{e^{-W}}{16\pi G r^2 {\cal F}^{1/2}}\left(\cot \th \pa_\th W-(\pa_\th W)^2+\pa^2_\th W \right)-\fr{1}{16\pi G r^2 {\cal F}^{3/2}}r\pa_v{\cal F}\left(2+r\pa_r W \right)\\
&+\fr{1}{16\pi G r^2 {\cal F}^{3/2}}\left(2r {\cal F}-r^2 \pa_r {\cal F} \right)\pa_v W-\fr1{8\pi G {\cal F}^{1/2}}\pa_v \pa_r W-\fr{e^W}{8\pi G {\cal F}^{3/2}}\pa^2_v W,\\
q_\th=&\fr{e^{-W}}{16\pi G {\cal F}^{1/2}}\left(\pa_r{\cal F}\pa_\th W+{\cal F}\left(-\pa_\th W \pa_r W+\pa_\th \pa_r W \right)\right)\\
&+\fr1{16\pi G {\cal F}^{1/2}}\left(2\cot\th+\pa_\th W \right)\pa_v W-\fr1{16\pi G {\cal F}^{1/2}}\pa_v \pa_\th W;
\end{split}
\la{qmVm}
\ee
\ssk
\[
\t_{mn}=\fr{1}{8\pi G}\left(
\begin{array}{cccc}
0&0&0&0\\
0& \tilde{\t}_{rr} & \tilde{\t}_{r\th} & 0\\
0& \tilde{\t}_{\th r} & \tilde{\t}_{\th\th}&0\\
0&0&0& \tilde{\t}_{\vf\vf}
\end{array}
\right),
\]
\be
\begin{split}
\tilde{\t}_{rr}=&\fr1{4r^2{\cal F}}\Bigg[ -4-4\cot\th \pa_\th W+(\pa_\th W)^2+4{\cal F}(1+r \pa_r W)+2 r \pa_r {\cal F}(2+r\pa_r W)\\
&-\fr{2e^W}{{\cal F}}r\pa_v{\cal F}\left(2+r\pa_r W \right) \Bigg]+\fr{e^W}{2r^2{\cal F}^2}\left(6r{\cal F}+r^2\pa_r{\cal F}+2r^2{\cal F}\pa_r{\cal F} \right)\pa_v W-\fr{e^{2W}}{{\cal F}}\pa^2_v W ,\\
\tilde{\t}_{r\th}=&\left(\fr1{r}+\fr{\pa_r{\cal F}}{2{\cal F}}\right)\pa_\th W+\cot\th \pa_r W+\fr{e^W}{2{\cal F}}\left[\left(2\cot\th+\pa_\th{\cal F} \right)\pa_v W-\pa_v\pa_\th W \right],
\\
\tilde{\t}_{\th\th}=&\fr{1}{4} \left[2r^2\pa^2_r{\cal F}+4\cot\th\pa_\th W+(\pa_\th W)^2-4r{\cal F}\pa_r W+2r\pa_r{\cal F}(2-r\pa_r W) \right]\\
&+e^W r^2\pa_v\pa_r W,
\\
\tilde{\t}_{\vf\vf}=&\fr14 e^{-2W}\sin^2\th \left[2r^2 \pa^2_r{\cal F}-(\pa_\th W)^2+4\pa^2_\th W+4r{\cal F}\pa_r W \right.\\
&\left.+2r\pa_r{\cal F}(2+r\pa_r W+4r^2{\cal F}\pa^2_r W) \right]\\
&+e^{-W}\sin^2\th \left[2\left(r+r^2\pa_r W \right)\pa_v W+3 r^2\pa_v \pa_r W \right].
\end{split}
\la{tauVm}
\ee

To check the energy conditions and to establish the viability of the proposed model  we have to set up the deformation function $W(v,r,\th)$. Again, as in the case of the deformed Schwarzschild solution in the null coordinates, we have not a chance to determine this function analytically, as a solution to the corresponding differential equations. So we propose to use the same arguments as in the previously considered case of the Schwarzschild-type solution in the null coordinates and choose this function as
\be
W(v,r,\th)=C_1 \ln \left(\fr{r_0 \sin \th}{r {\cal F}^{1/2}(v,r)}\right).
\la{fVm}
\ee 
For the sake of simplicity, we also fix $C_1=+1$.

Then, the NEC for the Vaidya-type metric \rf{Vm} with the deformation function \rf{fVm} turns out to be (we omit the insignificant coefficient $1/8\pi G$ on the r.h.s.)
\be
\begin{split}
\text{NEC}&=\fr{1}{4r^4_0\sin^4\th}\left(1-\fr{2M}{r} \right)^2 \left[r^2 \left(\fr{2}{\sin^2\th}-1 \right)+2M \left(r\left(1-\fr{2}{\sin^2\th} \right)-M \right) \right]\\
&+\fr{r(3r-4M)}{r^3_0 (r-2M) \sin^3\th} \left(1-\fr{2M}{r} \right)^{3/2}M^{\prime}\\
&-\fr{r}{r^2_0(r-2M)\sin^2\th}\left(2(M')^2+(r-2M)M^{\prime\prime} \right)\ge0. 
\end{split}
\la{NECVm}
\ee
In the case of a Vaidya-type metric it is difficult to make the analysis to the full extent. A simplification can be achieved for a very slowly variable mass function $M(v)$.\footnote{For example, the mass function, given in Ref. \cite{Song:2021ziq}, viz. $M(v)=M_0 (1+\tanh v)/2$, meets this criterion in a wide range of the function argument values.} However, even under these conditions, only qualitative analysis becomes possible. So, the NEC is achieved: 1) at $r\ge (1+\sqrt{3})M$, which is attainable in physical configuration of matter (see Ref. \cite{Song:2021ziq} in this respect); 2) for $M^{\prime} \equiv \pa_v M(v) >0$, which is a standard requirement for the Vaidya solution in ingoing coordinates; 3) and for $M^{\prime\prime} \equiv \pa^2_v M(v) <0$,\footnote{Explicitly, $2(M')^2+(r-2M)M^{\prime\prime}<0$ is required to provide the NEC.} which is a new requirement.

The energy density \rf{EVm} with the choice \rf{fVm} becomes
\be
\begin{split}
{\cal E}=&\fr{1}{64\pi G r^3(r-2M)^3}\Bigg[ \fr{r(r-2M)^3}{r^2_0 \sin^2\th}\left(\fr{r}{\sin^2\th} (11+3\cos 2\th)(r-2M)-8M^2 \right)\\
&+\fr{8r^3}{r_0 \sin\th} (r-2M)(3r-M)\sqrt{1-\fr{2M}{r}}\,M^\prime-8r^4 \left(2(M^\prime)^2+(r-2M) M^{\prime\prime} \right)\Bigg].
\end{split}
\la{WECVm}
\ee 
Here the positivity of the energy density is supported by the following conditions:
\be
{\cal E} \ge 0 \qquad \rightleftarrows \qquad r \ge (1+\sqrt{2})M,\,\,\,\,\, M^\prime > 0,\,\,\,\,\, M^{\prime\prime}<0.
\la{WECVmc}
\ee

For the fixed parameters of the solution the SEC (modulo $8\pi G$) is
\be
\begin{split}
\text{SEC}=\fr{1}{2r r^2_0 (r-2M)^3}\Bigg[(r-2M)^4\, \fr{\cos^2\th}{\sin^4 \th}&+\fr{2rr_0(r-2M)(r-5M)}{\sin\th}\sqrt{1-\fr{2M}{r}} M^\prime\\
&-2 r^2 r^2_0 \left(2(M^\prime)^2+(r-2M) M^{\prime\prime} \right) \Bigg]\ge 0 .
\end{split}
\la{SECVm}
\ee 
It is easy to see that the SEC can not be satisfied everywhere inside $r\ge 3M$.\footnote{Recall, $r\ge 3M$ is the location of equatorial circular orbits of photons in the previously considered stationary model. Since we analyze here the case of $|M^{\prime\prime}|\ll M^\prime \ll 1$, the location of these orbits in the non-stationary case is close to this value.} However, with the same restrictions on derivatives of the mass function as before (cf. eq. \rf{WECVmc}), the SEC is satisfied for $r \ge 5M$. In the preceding subsection, we have discussed reasons for having just massive particles in the matter content of the accretion disk for the corresponding stationary model. It becomes even more pronounced in the non-stationary case, since the requirement of having $r \ge 5M$ for the SEC fulfils for the ISCO of massive particles in the equatorial plane, for which $r \ge 6M$ is required.

The FEC is the most complicated condition to analyze. This condition reads (again, we omit the insignificant coefficient $8\pi G$)
\be
\begin{split}
&\text{FEC}=\fr{1}{64 r^3 r^4_0(r-2M)^6}\Bigg[\fr{16(r-2M)^3}{\sin^4\th}\times
\\
&\times\left((r-2M)^2(r+M)+3 r^2 r_0 \cos\th\sqrt{1-\fr{2M}{r}} M^\prime \right)^2\\
&+\fr{(r-2M)^2}{r\sin\th}\Psi^2-2\sqrt{1-\fr{2M}{r}} \Psi\left( \fr{8r r_0}{\sin\th}(r-2M)^2 (3r-M) M^\prime \right.\\
&\left.+\sqrt{1-\fr{2M}{r}} \left(\fr{64M^5}{\sin^2\th} +\fr{1}{\sin^4\th}\left(r^5(11+3\cos 2\th)-Mr^4 (88+24\cos 2\th) \right.\right.\right.\\
&\left.\left.\left.+M^2 r^3 (260+76\cos 2\th)-M^3 r^2 (328+120\cos 2\th)+M^4 r(128+96\cos 2\th)\right)\right.\right.\\
&\left.\left.-8r^3 r^2_0 \left(2(M^\prime)^2+(r-2M)M^{\prime\prime} \right)\right)\right)\Bigg]\le 0,
\end{split}
\la{FECVm}
\ee
where we have introduced
\be
\begin{split}
\Psi\equiv &\fr{r^4}{\sin^3\th} (7+3\cos 2\th)+4M\Bigg[\fr{10Mr^2}{\sin^3\th}(2+\cos 2\th)-\fr{20 M^2 r \cos^2\th}{\sin^3\th}-\fr{8M^3}{\sin \th}\\
&+3r^2\left(\fr{r}{\sin\th}\left(3-\fr{5}{\sin^2\th} \right)+2r_0 \sqrt{1-\fr{2M}{r}} M^\prime \right)\Bigg].
\end{split}
\la{PsidefVm}
\ee
Clearly, it is not so easy to make a comprehensive analysis of \rf{FECVm} (recall, this expression shall be negative to exclude a tachyonic-type matter). However, we can examine the FEC near the boundary value of circular equatorial orbits of massive particles. The series expansion of \rf{FECVm} at $r=6M$ and $\th=\pi/2$ reads
\be
\begin{split}
\text{FEC}\approx \fr{(44M+9\sqrt{6} r_0 M^\prime)}{2^8\cdot 3^4 r^4_0 M^3}\Big[-140 M^2-93\sqrt{6} r_0 M M^\prime+27 r^2_0 \left(2\left(M^\prime\right)^2+4M M^{\prime\prime}\right)\Big]\\
-\fr{1}{2^{10}\cdot 3^5 r^4_0 M^4}\Big[16000 M^3+6075\sqrt{6}r^3_0\left(M^\prime\right)^3+12 r_0 M^2 \left(-1045\sqrt{6} M^\prime+1008 r_0 M^{\prime\prime} \right)\\
+54 r^2_0 M M^\prime \left(-563 M^\prime+171\sqrt{6} r_0 M^{\prime\prime} \right)\Big]\left(r-6M \right)+\cO\left((r-6M)^2\right).
\end{split}
\la{FEC1Vm}
\ee
The series expansion \rf{FEC1Vm} becomes negative if 
\be
M^\prime >0,\qquad 2\left(M^\prime\right)^2+(r-2M)\Big|_{r=6M} M^{\prime\prime}<0,
\la{FEC2Vm}
\ee
which are the same requirements which have been found for the NEC, WEC and SEC.

To summarize this part of the paper, we have constructed various solutions for static (Schwarzschild-type) and non-stationary (Vaidya-type) non-rotating black holes with non-trivial external matter and have investigated their properties in terms of different energy conditions responsible for the physicality of matter.
In this connection, we note several interesting points. To begin with, we trace a curious feature of the characteristics of solutions depending on the choice of coordinates. For usual coordinates, which are the analog of Boyer-Lindquist coordinates in the Kerr solution, the stress-energy tensor of matter admits a representation as the stress-energy tensor of an anisotropic relativistic fluid. The field content of the accretion disk is not limited in any way: any physical fields, massive or massless, can be included in its composition. The situation drastically changes at transition to the null coordinates (Eddington-Finkelstein coordinates), when we have to consider the stress-energy tensor of matter fields as that of a viscous fluid. Using the same type of deformation function as for the stationary solution in usual coordinates leads to an interesting consequence: the field composition of the accretion disk should include only massive physical fields. Note that this restriction concerns only (conditionally) stable closed orbits of particles in the angular direction $\th=\pi/2$, and is not a general provision. The same constraints on the field composition of the accretion disk follow from the analysis of the energy conditions for the considered here non-stationary solution of the Vaidya-type black hole. 

Thus, we have established constraints on the parameters leading to the viability of solutions of different kinds for rotating and non-rotating black holes with matter fields around them. In the next section we extend our construction to the case of a simultaneously rotating and non-stationary black hole.

\section{Kerr-Vaidya-type black hole}

Having constructed the non-rotating stationary and non-stationary solutions with matter fields we have opened the way to constructing their rotating versions.  
Here we aim mainly at the principled construction of these solutions, postponing the analysis of the energy conditions for the corresponding stress-energy tensors to the future work.

First, let's construct a rotating version of \rf{SSm}. We will use the standard Newman-Janis algorithm \cite{Newman:1965tw} (see also \cite{Ghosh:2014hea,Dahal:2020hsj}) to this end. According to the algorithm, we have to present the inverse to \rf{SSm} metric in the Newman-Penrose formalism  \cite{Newman:1961qr}:
\be
g^{mn}=-l^m n^n-l^n n^m+m^m \bar{m}^n+m^n\bar{m}^m.
\la{invSSNP}
\ee
Here $l^m$ is one of two null-vectors (cf. eq. \rf{nullSSEF}); the other null-vector is
\be
n^m=\left(0,-1,0,0 \right),\qquad n^m n_m=0,\qquad l^m n_m=-1.
\la{ndefSSNP}
\ee
The complex conjugate vectors $m^m$ and $\bar{m}^m$, 
\be
m^m=\fr1{\sqrt{2}r}\left(e^{-W(r,\th)}\d^m_\th+\fr{i}{\sin \th}\d^m_\vf \right),\qquad \bar{m}^m=(m^m)^*,\qquad m^m \bar{m}_m=1,
\la{mdefSSNP}
\ee
are orthogonal to the null-vectors $l^m$ and $n^m$.

Next, following \cite{Newman:1965tw} one has to transform coordinates to
\be
x^{\prime\, m}=x^m-ia \left(\d^m_v+\d^m_r \right)\cos\th.
\la{xprimeNJ}
\ee
Treating the coordinates $v$ and $r$ as complex-valued ones, hence the real-valued functions $f(r)$ and $W(r,\th)$ from \rf{SSm} shall be considered as
\[
f\left(\fr12 (r+r^*)\right),\qquad  W \left(\fr12 (r+r^*),\th \right),
\]
we arrive at the following expressions for the transformed Newman-Penrose tetrad:
\be
\begin{split}
l^{\prime\,m}=&e^{-W(r,\th)}\d^m_v+\fr12 e^{-2W(r,\th)}\tilde{f}(r,\th)\d^m_r,\qquad n^{\prime\,m}=-\d^m_r, \\
m^{\prime\, m}=&\fr1{\sqrt{2}(r-ia\cos\th)}\left[e^{-W(r,\th)}\d^m_\th+ie^{-W(r,\th)}a \sin\th \left(\d^m_v+\d^m_r \right)+\fr{i}{\sin\th}{\d^m_\vf} \right],\\
\bar{m}^{\prime\, m}=&\fr1{\sqrt{2}(r+ia\cos\th)}\left[e^{-W(r,\th)}\d^m_\th-ie^{-W(r,\th)}a \sin\th \left(\d^m_v+\d^m_r \right)-\fr{i}{\sin\th}{\d^m_\vf} \right].
\end{split}
\la{NPprime}
\ee
The final step is to form the new inverse metric
\[
g^{\prime\,mn}=-l^{\prime\,m} n^{\prime\, n}-l^{\prime\, n} n^{\prime\, m}+m^{\prime\, m} \bar{m}^{\prime\, n}+m^{\prime\, n}\bar{m}^{\prime\, m},
\la{gprimeNJ}
\]
after inverting of which we finally get
\be
\begin{split}
ds^2=-\left(1-\fr{2M r}{\r^2} \right)dv^2+2e^{W(r,\th)} dv dr+e^{2W(r,\th)}\r^2 d\th^2\\
+2a\sin^2\th\left(-1+e^{-W(r,\th)}\left(1-\fr{2Mr}{\r^2} \right)\right) dv d\vf-2a\sin^2\th dr d\vf\\
+\sin^2\th \left(\r^2+ e^{-W(r,\th)}a^2\sin^2\th \left(2-e^{-W(r,\th)}\left(1-\fr{2Mr}{\r^2}\right)\right)\right)d\vf^2 .
\end{split}
\la{KerrgenNJ}
\ee
Here, as usual, $\r^2=r^2+a^2\cos^2\th$, and the presented in \rf{KerrgenNJ} metric is the Kerr-type generalization of the solution \rf{SSm}. 

Then the Kerr-Vaidya-type generalization of  \rf{KerrgenNJ} is a formalization of the construction for the case of ``time''-dependent black hole mass and ``time''-dependent deformation function. In its final form, the metric of the Kerr-Vaidya-type looks as follows:
\be
\begin{split}
ds^2=-\left(1-\fr{2M(v) r}{\r^2} \right)dv^2+2e^{W(v,r,\th)} dv dr+e^{2W(v,r,\th)}\r^2 d\th^2\\
+2a\sin^2\th\left(-1+e^{-W(v,r,\th)}\left(1-\fr{2M(v)r}{\r^2} \right)\right) dv d\vf-2a\sin^2\th dr d\vf\\
+\sin^2\th \left(\r^2+ e^{-W(v,r,\th)}a^2\sin^2\th \left(2-e^{-W(v,r,\th)}\left(1-\fr{2M(v)r}{\r^2}\right)\right)\right)d\vf^2 .
\end{split}
\la{KerrVgen}
\ee
Apparently, for the trivial deformation function $W(v,r,\th)$ the line element \rf{KerrVgen} turns into the well-known metric of the Kerr-Vaidya black hole (cf. Ref. \cite{Dahal:2020hsj}).

As well as in the case of the generalized Vaidya black hole model, the solution \rf{KerrVgen} corresponds to the configuration of a rotating black hole surrounded by matter fields. A detailed study of the energy conditions on matter fields is beyond the scope of the present paper. However, even on the qualitative level with taking into account of the results already obtained, some conclusions can be drawn. For instance, the stress-energy tensor of matter for this type solution will have the form of the full stress-energy tensor of a viscous fluid. Further, the deformation function for \rf{KerrVgen} should be a generalization of \rf{Solgen}. However, this generalization will be more concerned with the dependence on time-like coordinate, leaving the angular dependence unchanged. Therefore, extra singularities in the pole (north and south) directions can be expected to be preserved to some extent. As an additional argument in favor of this claim we refer to Ref. \cite{Nurmagambetov:2020ann}, where the bipolar outflow in the model of time-depending rotating black hole with matter had been also established, but in a different context.

\section{Discussion of the results and conclusions}

To sum up, in this paper we have outlined the strategy in constructing different solutions to the Einstein equations for rotating and non-rotating black holes in the environment of matter fields. This type of solutions can be considered as an initial or intermediate stage of black hole formation, the final result of which is either complete absorption of matter by the black hole or creation of a quasi-stable configuration of the black hole with an accretion disk of external matter. If we are supposing a complete absorption of matter in the process of formation of a rotating black hole, then at construction of new solutions it is necessary to take into account the symmetry properties of the Kerr solution as a final stage of such process. Retaining the symmetry and singularity properties of the standard Kerr solution as much as possible restricts the form of the deformation function generating a non-trivial effective stress-energy tensor of matter. However, the viability of solutions with a nontrivial stress-energy tensor is determined by various energy conditions, to which this tensor has to satisfy. Therefore, the fulfillment of this requirement is also important for further fixing the explicit form of the deformation function. 

Verifying the singularities of the constructed metric we have observed the important difference with respect to the standard Kerr solution. Specifically, in addition to the singularity of the Kerr solution, there are other points of singularity in our model, located in the north and south pole directions. In dependence on the choice of the deformations function, this additional singularity of the metric can be pronounced or smoothed by the exponential suppression (cf. eqs.  \rf{K+1} and \rf{Kl-2}). But even in the case when the additional singularity is smoothed, the exponential suppression in the corresponding directions results in a near-zero value of the Kretschmann scalar. Gravitational field becomes weak at the north and south poles, and particles from the black hole ergoregion may freely escape the black hole, forming a well-known picture of the bipolar outflow.

To fix the free parameters of the considered metric we have applied various popular energy conditions on the stress-energy tensor of matter required for the viability of our model. We observed that the most of the energy conditions, including the weak and strong energy conditions important in astrophysics, can be satisfied, so that the matter fields of the accretion disk around the black hole do not suffer from pathologies. As a by-product of our studies, we have verified the viability of the construction in the case of (A)dS space.

Having established the viable solutions for stationary rotating black holes in the matter fields environment, we have proceeded with extending the approach to the non-stationary case of the Kerr-Vaidya black holes. Various solutions for static (Schwarzschild-type) and non-stationary (Vaidya-type) non-rotating black holes with non-trivial external matter have been constructed on this route, and their properties in terms of different energy conditions providing the physicality of matter have been investigated. And here we have observed several interesting points. Specifically, we figured out a curious feature of the characteristics of solutions depending on the choice of coordinates. For the Boyer-Lindquist type coordinates the stress-energy tensor of matter admits a representation of the  stress-energy tensor of an anisotropic relativistic fluid. The field content of the accretion disk is not limited in any way: any physical fields, massive or massless, can be included in its composition. The situation changes upon the transition to the null coordinates (Eddington-Finkelstein coordinates), when the stress-energy tensor of matter becomes the complete tensor of a viscous fluid. The use of the same deformation function as for the solution in the Boyer-Lindquist coordinates leads to an unexpected consequence: the field composition of the accretion disk should include only massive physical fields. Note that this restriction concerns only (conditionally) stable closed orbits of particles in the equatorial plane, and is not a general provision. The same constraints on the field composition of the accretion disk follow from the analysis of the energy conditions for the found non-stationary solution of the Vaidya-type black hole.

Finally, on the ground of the obtained results for the generalized Vaidya-type black hole, we have extended the construction to the case of a rotating Kerr-Vaidya-type solution in its general form, by use of the standard algorithm by Newman and Janis. We have performed a brief qualitative analysis of the solution properties and will postpone a full analysis to future publications. We hope that our findings will be useful in studies of astrophysical black holes in a broad context, especially in the case, when the accretion disk of matter becomes an indispensable attribute.

\bsk
{\bf Acknowledgements}. 
The author would like to thank Vira and Vasyl Oryol for their help and kind hospitality.
The work is supported in part within the Cambridge-NRFU 2022 initiative ''Individual research (developments) grants for researchers in Ukraine (supported by the University of Cambridge, UK)'', project №2022.02/0052.


\bsk
{\bf Conflict of Interest}.
The author declare no conflict of interest.

\newpage
\appendix
\numberwithin{equation}{section}

\section{The Kretschmann scalar for the deformed solution}
\la{app:F-exact}

For the line element \rf{Rotellipsm} the r.h.s. of the following expression for the Kretschmann scalar (eq. \rf{KKerrgen} in the main text),
\[
K=e^{-4 W(r,\th)}K_{\text{Kerr}}+F(W(r,\th), r,\th;M,a),
\]
includes the function $F(W(r,\th), r,\th;M,a)$. The explicit form of this function comes as follows:
\be
\begin{split}
&F(W(r,\th), r,\th;M,a)=\fr{4 e^{-4W(r,\th)}}{\r^{12}}\times\\
&\times \Bigg[ \left(\left(D_\th W(r,\th)\right)^2 +\D^2 \left(\pa^2_r W(r,\th)\right)^2\right) \r^8+2Mr \D \S \r^4 \pa^2_r W(r,\th)\\
&+2 D_\th W(r,\th)\left[\left((r-M)\pa^2_r W(r,\th)+\D \pa^2_r W(r,\th)\right) \r^4+Mr\S \right]\r^4 \\
&+2\r^2\cot\th\, \pa_\th W(r,\th)\Big[\D\pa^2_r W(r,\th)\left( r^6+a^6\cos^6\th+3\r^2 r^2 a^2 \cos^2\th \right)\\
&+\r^6\left((r-M)\pa_r W(r,\th)+D_\th W(r,\th)\right)\\
&+4Mr \left(r^4+6r^2 a^2-6 a^4\cos^2\th-8 r^2 a^2\cos^2\th+3 a^4 \cos^4\th \right)\Big]+2\r^2 \pa_r W(r,\th)\times \\
&\times \Big[2M\left(-3a^6\cos^4\th +3ra^4 \left(4M\cos^2\th-3r(\cos^2\th-2)\right)\cos^2\th \right.\\
&
\left. +r^3 a^2 \left(r(14\cos^2\th-3)-32 M\cos^2\th\right)+(4M-r)r^5\right)+(r-M)\D\r^6\pa^2_r W(r,\th)\Big] 
\\
&+\fr{\r^4}{\D\sin^2\th}\left(\pa_\th W(r,\th)\right)^2\times \\
&\times \Big[3a^6\cos^6\th+r^4 \left(10M^2\sin^2\th+2Mr(\cos^2-4)+r^2(\cos^2\th+2)\right)
\\
&+a^4\left(2M^2\sin^2\th \cos^4\th+r^2(\cos^2\th+8)-2Mr(\cos^4\th+8\cos^2\th-6)\cos^2\th \right)\\
&+r^2a^2\left(-20M^2\sin^2\cos^2+r^2(2\cos^2\th+7)\cos^2\th-4Mr(4\cos^4\th-2\cos^2\th+1)\right)\Big]\\
&-\fr{\r^4}{\sin^2\th}\left(\pa_r W(r,\th)\right)^2 \times\\
&\times \Big[-2a^6\cos^6\th+r^4\left(-11M^2 \sin^2\th+2Mr(5-3\cos^2\th)+r^2(\cos^2\th-3)\right)\\
&+a^4\left(r^2(\cos^2-7)\cos^4\th-3M^2\sin^2\th\cos^4\th-2Mr(\cos^4\th-9\cos^2\th+6)\cos^2\th \right)\\
&+2r^2 a^2 \left(r^2(\cos^2\th-4)\cos^2\th+9M^2\sin^2\th\cos^2\th+2Mr(2\cos^4\th-\cos^2\th+1)\right)\Big]\Bigg].
\end{split}
\la{Fdef}
\ee
To simplify the expressions a bit,  the following quantities have been used in \rf{Fdef}:
\be
D_\th W(r,\th)\equiv \pa^2_\th W(r,\th)-\cot \th \pa_\th W(r,\th) ; \quad \S\equiv \r^2-4a^2 \cos^2\th;
\la{D2thdef}
\ee

From \rf{Fdef} it is clear that the symmetry under $\th \ra -\th$ of the Kerr solution can be preserved with the appropriate choice of the deformation function $W(r,\th)$. Also, the appeared in \rf{Fdef} additional potential point of singularity at the horizon $\D=0$ can be avoided by the suitable choice of $W(r,\th)$.

\section{Kretschmann scalars: Details}
\la{app:Kretsch-details}

In this appendix, we give explicit expressions for the Kretschmann scalars from Section 2.3 of the main text.

We begin with eq. \rf{K-1}. Here we have
\[
K_{\big|_{C_1=-1}}=\fr{4r^4_0}{\D^4 \r^{12}}
 \Bigg[3 a^8 \left(a^2-M^2\right)^2 \cos^{12}\theta +r^6 \left[4 a^4 \left(6 M^2-2 M r+r^2\right) \right.
\]
\[
\left.+4 a^2 r \left(-24 M^3+19 M^2 r-6 M r^2+r^3\right) \right.
\]
\[
\left. +r^2 \left(99 M^4-132 M^3 r+70 M^2 r^2-20 M r^3+3 r^4\right)\right]
\]
\[
-2 a^6 \left(a^2-M^2\right) \cos^{10}\theta  \left(2 a^4+a^2 \left(-3 M^2+2 M r-7 r^2\right)+6 M^2 r^2\right) 
\]
\[
+2 r^4 \cos^2 \theta \left[a^6 \left(-120 M^2+4 M r+8 r^2\right)+2 a^4 r \left(240 M^3-144 M^2 r-2 M r^2+3 r^3\right) \right.
\]
\[
\left. +a^2 r^2 \left(-474 M^4+584 M^3 r-167 M^2 r^2-14 M r^3+7 r^4\right)
\right.
\]
\[
\left. +M r^4 \left(-99 M^3+114 M^2 r-41 M r^2+4 r^3\right)\right]
\]
\[
+a^4 \cos^8\theta \left[4 a^8-4 a^6 \left(M^2+8 M r+3 r^2\right)+a^4 \left(3 M^4-4 M^3 r+270 M^2 r^2-76 M r^3+29 r^4\right)\right.
\]
\[
\left. -4 a^2 M r^2 \left(6 M^3+140 M^2 r-73 M r^2+6 r^3\right)+6 M^2 r^4 \left(83 M^2-80 M r+20 r^2\right)\right]
\]
\[
+r^2 \cos^4\theta  \left[8 a^8 \left(15 M^2+5 M r+3 r^2\right)+8 a^6 r \left(-60 M^3+67 M^2 r+3 M r^2+r^3\right)\right.
\]
\[
\left. +a^4 r^2 \left(498 M^4-2264 M^3 r+1100 M^2 r^2-64 M r^3+29 r^4\right) \right.
\]
\[
\left. +2 a^2 M r^4 \left(948 M^3-1016 M^2 r+273 M r^2-4 r^3\right) +3 M^2 r^6 \left(33 M^2-32 M r+8 r^2\right)\right]
\]
\[
+4 a^2 r \cos^6\theta \left(a^8 (6 M+4 r)-2 a^6 r \left(44 M^2+3 M r+r^2\right) \right.
\]
\[
\left. +a^4 r \left(3 M^4+260 M^3 r-135 M^2 r^2-32 M r^3+9 r^4\right) \right.
\]
\be
\left. -a^2 M r^3 \left(249 M^3-446 M^2 r+149 M r^2+10 r^3\right)-3 M^2 r^5 \left(79 M^2-80 M r+20 r^2\right)\right)\Bigg].
\la{K-1-expl}
\ee

Next, the explicit form of the Kretsch\-mann scalar from eq. \rf{K+1} is
\[
K_{\big|_{C_1=+1}}=\fr{4}{r^4_0\r^{12}\sin^8\th} \Bigg[3 a^8 \left(a^2-M^2\right)^2 \cos^{12}\theta 
\]
\[
+r^7 \left(4 a^4 (2 M+r)-4 a^2 \left(2 M^3+5 M^2 r-r^3\right) \right.
\]
\[
\left. +r \left(19 M^4-12 M^3 r+14 M^2 r^2-12 M r^3+3 r^4\right)\right)
\]
\[
-2 a^6 \cos^{10}\theta  \left(2 a^6+7 a^4 \left(M^2+2 M r-r^2\right)+a^2 M \left(3 M^3-62 M^2 r+13 M r^2+12 r^3\right)\right.
\]
\[
\left.+6 M^2 r^2 \left(11 M^2-10 M r+2 r^2\right)\right)
\]
\[
-2 r^4 \cos^2\theta  \left(4 a^6 \left(15 M^2+M r-2 r^2\right)+2 a^4 r \left(-122 M^3+60 M^2 r+14 M r^2-3 r^3\right)\right.
\]
\[
\left.+a^2 r^2 \left(242 M^4-244 M^3 r+11 M^2 r^2+34 M r^3-7 r^4\right)+M^2 r^4 \left(19 M^2-10 M r+r^2\right)\right)
\]
\[
+a^4 \cos^8\theta  \left(4 a^8+4 a^6 \left(11 M^2+10 M r-3 r^2\right) \right.
\]
\[
\left. +a^4 \left(3 M^4-244 M^3 r+294 M^2 r^2-20 M r^3+29 r^4\right) \right.
\]
\[
\left.+4 a^2 M r^2 \left(66 M^3-286 M^2 r+141 M r^2-10 r^3\right)+2 M^2 r^4 \left(449 M^2-460 M r+120 r^2\right)\right)
\]
\[
-4 a^2 \cos^6\theta  \left(a^8 \left(6 M^2+6 M r-4 r^2\right)+2 a^6 r \left(-15 M^3+62 M^2 r-3 M r^2+r^3\right)\right.
\]
\[
\left.+a^4 r^2 \left(33 M^4-486 M^3 r+287 M^2 r^2-9 r^4\right)\right.
\]
\[
\left. +a^2 M r^4 \left(449 M^3-588 M^2 r+185 M r^2+2 r^3\right)+M^2 r^6 \left(121 M^2-122 M r+30 r^2\right)\right)
\]
\[
+r^2 \cos^4\theta  \left(8 a^8 \left(30 M^2-5 M r+3 r^2\right)+8 a^6 r \left(-115 M^3+97 M^2 r-9 M r^2+r^3\right)\right.
\]
\[
\left.+a^4 r^2 \left(898 M^4-1920 M^3 r+812 M^2 r^2-64 M r^3+29 r^4\right)\right.
\]
\be
\left.+2 a^2 M r^4 \left(484 M^3-484 M^2 r+105 M r^2+4 r^3\right)+M^3 r^6 (19 M-8 r)\right)\Bigg].
\la{K+1-expl}
\ee

The explicit form of eq. \rf{Kl>0gen} can be obtained by use of eqs. \rf{KKerr}, \rf{KKerrgen}, \rf{f2}, \rf{Fdef} and \rf{D2thdef}, the Kretschmann scalar becomes in this case equal to
\[
K^{(l)}_{C_2}=\frac{4 e^{-4 C_2 \D^{-\frac{l}{2}} \sin ^l \theta }}{\r^{12}} \Bigg[2 C_2 l\r^4 \left((l-1) \cos^2\th -1\right)  \sin^{l-2}\th \times
\]
\[
\times \D^{-\frac{l}{2}} \left(C_2 l  \r^4 \D^{-\frac{l}{2}-1} \left((l+1)\D+(l+2)( M^2-a^2)-(M-r)^2\right)\sin^{l}\th+M r \S\right)
\]
\[
+2 C_2 l M r \sin^{l}\th \r^4 \S \D^{-\frac{l}{2}-1} \left((l+1)\D+(l+2)( M^2-a^2)\right)
\]
\[
-2C_2( r- M)\sin^{l}\th \r^2 \D^{-\frac{l}{2}-1}\left(2 M \left(-3 a^6 \cos^4 \th+3 a^4 r \cos^2\th \left(4 M \cos^2\th-3 r \left(\cos^2\th-2\right)\right)\right.\right.
\]
\[
\left.\left.+a^2 r^3 \left(r \left(14 \cos^2\th-3\right)-32 M \cos^2\th\right)+r^5 (4 M-r)\right)\right.
\]
\[
\left.-C_2 l \sin^{l}\th (M-r) \r^6 \left((l+1)\D+(l+2)( M^2-a^2)\right)  \D^{-\frac{l}{2}-1}\right)
\]
\[
+C_2^2 l^2 \sin^{2l}\th \r^8 \left((l+1)\D+(l+2)( M^2-a^2)\right)^2 \D^{-l-2}
\]
\[
-C_2^2 l^2 \sin^{2l-2}\th (M-r)^2 \r^4 \D^{-l-2} \left(-2 a^6 \cos^6\th+a^4 \left(-3 M^2 \sin^2\th \cos^4\th \right.\right.
\]
\[
\left.\left. -2 M r \left(\cos^4\th-9 \cos^2\th+6\right) \cos^2\th +r^2 \left(\cos^2\th-7\right) \cos^4\th\right)\right.
\]
\[
\left. +2 a^2 r^2 \left(9 M^2 \sin^2\th \cos^2\th+2 M r \left(2 \cos^4\th-\cos^2\th+1\right)+r^2 \left(\cos^2\th-4\right) \cos^2\th\right)\right.
\]
\[
\left.+r^4 \left(-11 M^2 \sin^2\th+2 M r \left(5-3 \cos^2\th\right)+r^2 \left(\cos^2\th-3\right)\right)\right)
\]
\[
+C_2^2 l^2 \cos^2\th \sin^{2l-4}\th \r^4 \D^{-l-1} \left(3 a^6 \cos^6\th+a^4 \left(2 M^2 \sin^2 \th \cos^4\th \right.\right.
\]
\[
\left.\left. -2 M r \left(\cos^4\th+8 \cos^2\th-6\right) \cos^2\th +r^2 \left(\cos^2\th+8\right) \cos^4\th\right)
\right.
\]
\[
\left.
+a^2 r^2 \left(-20 M^2 \sin^2\th \cos^2\th-4 M r \left(4 \cos^4\th-2 \cos^2\th+1\right)+r^2 \left(2 \cos^2\th+7\right) \cos^2\th\right)\right.
\]
\[
\left.+r^4 \left(10 M^2 \sin^2\th+2 M r \left(\cos^2\th-4\right)+r^2 \left(\cos^2\th+2\right)\right)\right)
\]
\[
+2 C_2 l \cos^2\th \sin^{l-4}\th \r^2 \left(C_2 l a^8\left(l \cos^2\th-2\right)   \cos^6\th \sin^{l}\th \right.
\]
\[
\left. +a^2 r^4\left(C_2 l r^2\left(-3 \cos^4\th+(4 l-3) \cos^2\th-2\right)  \sin^{l}\th \right.\right.
\]
\[
\left.\left.+M^2 \sin^2\th \left(\cos^2\th \left(64 \D^{l/2}+3C_2 l (l+1)  \sin^{l}\th\right)-48 \D^{l/2}\right) \right.\right.
\]
\[
\left. \left. +2 M r \left(14 \D^{l/2}-3 \cos^2\th \left(10 \D^{l/2}+C_2 l(l-1)  \sin^{l}\th\right) +\cos^4\th \left(16 \D^{l/2}+3 C_2 l \sin^l \th \right)\right)\right)  \right.
\]
\[
\left.+a^4 r^2\left(3 C_2 l r^2\left(-\cos^4\th+(2 l-1) \cos^2\th-2\right)  \cos^2\th \sin^{l}\th \right.\right.
\]
\[
\left.\left. -3 \sin^2\th M^2 \left(\cos^2\th \left(8 \D^{l/2}-C_2 l (l+1) \sin^{l}\th\right)-16 \D^{l/2}\right) \cos^2\th \right.\right.
\]
\[
\left.\left. -2 M r \left(40 \cos^2\th \D^{l/2}-12 \D^{l/2}+\left(3 C_2 (l-1) l \sin^{l}\th-34 \D^{l/2}\right) \cos^4\th \right.\right.\right.
\]
\[
\left.\left.\left.+\cos^6\th \left(6 \D^{l/2}-3 C_2 l \sin^{l}\th\right)\right)\right) -a^6 \cos^2\th \left(-C_2 l (l+1) M^2 \cos^4\th \sin^{l+2}\th \right.\right.
\]
\[
\left.\left.+
C_2 l r^2\left(\cos^4\th-(4 l-1) \cos^2\th+6\right)  \cos^2\th \sin^{l}\th \right.\right.
\]
\[
\left.\left. +2 M r \left(-18 \cos^2\th \D^{l/2}+12 \D^{l/2}-C_2 l \cos^6\th \sin^l \th \right.\right.\right.
\]
\[
\left.\left.\left.+\cos^4\th \left(6 \D^{l/2}+C_2 l (l-1)  \sin^{l}\th\right)\right)\right)\right.
\]
\[
\left.+r^6 \D^{-l-1} \left(C_2 l r^2\left(-\cos^2\th+l-1\right)  \sin^l \th-\sin^2\th M^2 \left(8 \D^{l/2}-C_2 l (l+1) \sin^{l}\th\right)\right.\right.
\]
\[
\left.\left.-2 M r \left(-2 \D^{l/2}+C_2 l (l-1) \sin^{l}\th+\cos^2\th \left(2 \D^{l/2}-C_2 l \sin^{l}\th\right)\right)\right)\right) 
\]
\[
\left.+C_2^2 l^2 \left((l-1) \cos^2\th-1\right)^2 \sin^{2l-4}\th \r^8 \D^{-l} \right.
\]
\be
\left.
+12 M^2 \left(-a^6 \cos^6\th+15 a^4 r^2 \cos^4\th-15 a^2 r^4 \cos^2\th+r^6\right)\right)\Bigg].
\la{KC2gen}
\ee
Recall, $\Sigma$ has been determined in eq. \rf{D2thdef}.

Finally, we reproduce eq. \rf{Kl-2} in its full extent:
\[
K_{l=-2}=\frac{16 e^{-\frac{4 \D }{r_0^2 \sin ^2 \theta}}}{r_0^4\r^{12}\sin^8\th} \Bigg[-3 a^6 M^2 r_0^4 \cos^{14}\th + 3 a^4 \bigg(a^8 - 2 M^2 a^6 + M \left(M^3 - 2 r_0^2 M - 4 r r_0^2 \right) a^4 
\]
\[
+ 2 M^2 r_0^2 \left(-r^2 + 4 M r + 2 r_0^2\right) a^2 + 15 M^2 r^2 r_0^4 \bigg) \cos^{12}\th
\]
\[
+a^2 \bigg(4 a^{10} + 2 \left(M^2 - 16 r M + 11 r^2\right) a^8 
\]
\[
- 2 M \left(3 M^3 - 16 r M^2 + 5 r^2 M - 9 r_0^2 M + 6 r^3 - 18 r r_0^2\right) a^6
\]
\[
-2 M \left(6 r^2 M^3 - 6 (r^3 - 6 r r_0^2) M^2 + (9 r_0^4 - 24 r^2 r_0^2) M + 10 r^3 r_0^2\right) a^4 
\]
\[
- 10 M^2 r^2 r_0^2 \left(-3 r^2 + 4 M r + 18 r_0^2\right) a^2 - 45 M^2 r^4 r_0^4 \bigg) \cos^{10}\th
\]
\[
+\bigg(4 a^{12} + 4 \left(M^2 - 3 r M + 7 r^2\right) a^{10} 
\]
\[
+ \left(3 M^4 - 44 r M^3 + 2 (31 r^2 - 9 r_0^2) M^2 - 36 r (4 r^2 + r_0^2) M + 69 r^4 \right) a^8
\]
\[
+4 M \left(-8 r^5 + 15 r_0^2 r^3 + 6 M^3 r^2 + M^2 (7 r^3 + 18 r_0^2 r) + 3 M (r^4 - 9 r_0^2 r^2 + r_0^4)\right) a^6
\]
\[
+2 M r^2 \left(-11 r^2 M^3 + 10 (r^3 + 6 r_0^2 r) M^2 - 15 (2 r^2 r_0^2 - 9 r_0^4) M - 2 r^3 r_0^2\right) a^4
\]
\[
+2 M^2 r^4 r_0^2 \left(15 (r^2 + 6 r_0^2) - 28 M r \right) a^2 + 3 M^2 r^6 r_0^4 \bigg) \cos^8\th
\]
\[
+\bigg(16 r^2 a^{10} + 2 \left(36 r^4 + 6 M^3 r + M^2 (3 r_0^2 - 4 r^2) + M (6 r r_0^2 - 32 r^3)\right) a^8
\]
\[
-\left(-116 r^6 + 12 M^4 r^2 + 12 M^3 (5 r^3 + 2 r_0^2 r) + 12 M (24 r^5 + 5 r_0^2 r^3) \right.
\]
\[
\left. - 3 M^2 (60 r^4 + 32 r_0^2 r^2 - r_0^4)\right) a^6
\]
\[
+4 M r^2 \left(-6 r^5 + 3 r_0^2 r^3 + 11 M^3 r^2 + M^2 (r^3 - 30 r r_0^2) + M (r^4 - 45 r_0^4)\right) a^4
\]
\[
+2 M r^4 \left(2 r^2 M^3 + 2 (r^3 + 42 r_0^2 r) M^2 - 3 (45 r_0^4 + 16 r^2 r_0^2) M + 2 r^3 r_0^2\right) a^2
\]
\[
+2 M^2 r^6 r_0^2 \left(4 M r - 3 (r^2 + 2 r_0^2)\right) \bigg) \cos^6\th
\]
\[
+r^2 \bigg(24 r^2 a^8 + 2 \left(44 r^4 + 10 M^3 r + 10 M (r_0^2 - 6 r^2) r - M^2 (8 r^2 + 15 r_0^2)\right) a^6
\]
\[
+\left(109 r^6 - 22 M^4 r^2 + M^3 (40 r r_0^2 - 28 r^3) - 4 M (80 r^5 + 3 r_0^2 r^3) \right. 
\]
\[
\left. + 3 M^2 (84 r^4 + 20 r_0^2 r^2 + 15 r_0^4)\right) a^4
\]
\[
-2 M r^2 \left(4 r^2 M^3 + (84 r r_0^2 - 26 r^3) M^2 + (19 r^4 - 54 r_0^2 r^2 - 90 r_0^4) M + 6 r^3 r_0^2 \right) a^2
\]
\[
+M^2 r^4 \left(11 M^2 r^2 + 18 r_0^2 (r^2 + r_0^2) - 4 M (r^3 + 6 r_0^2 r)\right)\bigg) \cos^4\th
\]
\[
+r^4 \bigg(16 r^2 a^6 + \left(52 r^4 + 4 M^3 r + M (4 r r_0^2 - 96 r^3) + M^2 (8 r^2 - 30 r_0^2)\right) a^4
\]
\[
+\left(54 r^6 + 12 M (r_0^2 - 16 r^2) r^3 + 4 M^4 r^2 + M^3 (56 r r_0^2 - 52 r^3) \right.
\]
\[
\left. + M^2 (202 r^4 - 48 r_0^2 r^2 - 45 r_0^4)\right) a^2
\]
\[
-2 M r^2 \left(-2 r^5 + 11 M^3 r^2 - 2 M^2 (11 r^3 + 6 r_0^2 r) + M (13 r^4 + 9 r_0^2 r^2 + 6 r_0^4)\right)\bigg) \cos^2\th
\]
\[
+r^6 \bigg(11 r^6 - 48 M r^5 + 4 a^4 r^2 + 11 M^4 r^2 - 8 M^3 (5 r^2 + r_0^2) r 
\]
\[
+ M^2 (70 r^4 + 6 r_0^2 r^2 + 3 r_0^4)
\]
\be
-2 a^2 \left(-6 r^4 + 2 M^3 r + 2 M (7 r^2 + r_0^2) r -3 M^2 (2 r^2 + r_0^2)\right)\bigg)\Bigg] .
\la{K-2-expl}
\ee

\section{Stress-energy tensor in flat and (A)dS spacetimes}
\la{app:deformed-Kerr-SET}

Direct calculations of the Einstein tensor $G_{mn}\equiv R_{mn}-1/2 g_{mn} R$ for the metric \rf{Rotellipsm} in Minkowski spacetime give, by means of the Einstein equations $G_{mn}=8\pi T_{mn}$, the following non-trivial components of the stress-energy tensor $T_{mn}$:
\be
8\pi T_{tt}=-\fr{e^{-2W}(\r^2-2Mr)}{\r^4}\left[\D \,\pa^2_r W+(r-M) \pa_r W+\pa^2_\th W \right],
\la{G11}
\ee
\be
8\pi T_{t\vf}=-\fr{2e^{-2W}Mr a \sin^2\th}{\r^4}\left[\D \,\pa^2_r W+(r-M) \pa_r W+\pa^2_\th W   \right],
\la{G14}
\ee
\be
8\pi T_{rr}=\fr1{\D}\left[(r-M)\pa_r W-\cot\th \,\pa_\th W \right],
\la{G22}
\ee
\be
8\pi T_{r\th}=-\fr1{\sin\th}\left[\fr{r-M}{\D} \pa_\th W+\cot \th \,\pa_r W \right],
\la{G23}
\ee
\be
8\pi T_{\th\th}=-\fr{1}{\sin^2\th}\left[(r-M)\pa_r W-\cot\th \,\pa_\th W \right],
\la{G33}
\ee
\be
8\pi T_{\vf\vf}=\fr{e^{-2W}\sin^2\th}{\r^4}\left(\D \, \r^2 +2Mr(r^2+a^2) \right)
\left[\D \,\pa^2_r W+(r-M)\pa_r W+\pa^2_\th W \right].
\la{G44}
\ee

For the (A)dS space with metric \rf{KerrAdSm},
\[
8 \pi G T_{tt}=\fr{e^{-2W}}{2L^2\r^4}\left(\D_r-\D_\th a^2\sin^2\th \right)\times
\]
\be
\times\left[ 6\left(e^{2W}-1 \right)\r^2-2a^2 \pa_\th W\sin\th\cos\th -2L^2\D_\th  \pa_\th^2 W-2\Phi\pa_r W -2L^2\D_r  \pa^2_r W \right],
\la{T11AdS}
\ee
\[
8 \pi G T_{t\vf}=\fr{ae^{-2W} \sin^2\th}{\left(L^2-a^2\right)\r^4} \left(\D_r -(r^2+a^2) \D_\th  \right)\times
\]
\be
\times\left[ 3\left(1-e^{2W} \right)\r^2+a^2\sin\th \cos \th\pa_\th W+L^2\D_\th  \pa^2_\th W+\Phi \pa_r W+L^2 \D_r \pa^2_r W\right],
\la{T14AdS}
\ee
\be
8 \pi G T_{rr}=\fr1{\D_r L^2} \left[3\left(-1+e^{2W} \right)\r^2-\D_\th L^2 \cot\th \pa_\th W+\Phi \pa_r W \right],
\la{T22AdS}
\ee
\be
8 \pi G T_{r\th}=-\fr1{\D_\th L^2\sin\th} \left[\fr{\Phi \D_\th}{\D_r } \pa_\th W+\left(\D_\th L^2+a^2\sin^2\th \right)\cot\th \pa_r W \right],
\la{T23AdS}
\ee
\be
8 \pi G T_{\th\th}=-\fr1{\D_\th L^2 \sin^2\th}\left[3\left(-1+e^{2W} \right)\r^2-\D_\th L^2 \cot\th \pa_\th W+\Phi \pa_r W \right],
\la{T33AdS}
\ee
\[
8 \pi G T_{\vf\vf}=-\fr{\sin^2\th}{2\left(L^2-a^2\right)^2\r^4}\left[e^{-2W}L^2\left(\D_r  a^2\sin^2\th-(r^2+a^2)^2 \D_\th  \right) \left(2\Phi \pa_r W+2\D_r L^2 \pa^2_r W \right.\right.
\]
\be
\left.\left.+2a^2 \sin\th \cos\th \pa_\th W+2\D_\th L^2 \pa^2_\th W+6\r^2 \right)-6\r^2 L^2 \left(\D_r a^2 \sin^2\th-(r^2+a^2)^2 \D_\th \right)\right] .
\la{T44AdS}
\ee
Here we have introduced $\Phi\equiv L^2(r-M)+r(2r^2+a^2)$. A brief inspection shows that expressions \rf{T11AdS}--\rf{T44AdS} turn into their flat space counterparts \rf{G11}--\rf{G44} in the limit $L\ra \infty$.

\section{The zero cosmological limit of the (A)dS deformation function}
\la{app:Zero-lambda-W}

Let's outline the subtleties in taking the zero cosmological constant limit of eq. \rf{SolgenAdS}, which should lead to eq. \rf{Solgen}. The part of eq. \rf{SolgenAdS} with $C_2$ integration constant can be presented as
\be
\left(\fr{\sin^2\th \D_\th}{\D_r} \right)^{\fr1{2} L^2 l}=\left(\fr{\sin^l\th \D_\th}{\D_r^{l/2}}\right)^{L^2}.
\la{C2mod1}
\ee
Hence we have to analyze the following expression:
\be
\lim_{L^2\ra \infty} \left(\fr{\sin^l\th \D_\th}{\D_r^{l/2}}\right)^{L^2}=\lim_{L^2\ra \infty}\left(\fr{\sin^l\th }{\D^{l/2}}\right)^{L^2}=\left(\fr{\sin^l\th }{\D^{l/2}}\right) \lim_{L^2\ra \infty}\left(\fr{\sin^l\th }{\D^{l/2}}\right)^{L^2-1}.
\la{C2mod2}
\ee
The $L^2 \ra \infty$ limit is equivalent to $\lim_{n\ra \infty} x^{-n}$, where $n=L^2-1 \approx L^2$, and $x=\D^{l/2}/\sin^l \th$. In what follows we will consider the case of positive $x$, that turns us back to the even valued $l$s, as in the flat spacetime case. Then,
\be
\lim_{n\ra \infty} x^{-n}=\lim_{n\ra \infty} \ln e^{x^{-n}}\approx \lim_{n\ra \infty} \ln \left(1-x^{-n} \right)\sim\lim_{\ve \ra 0} \ve.
\la{C2mod3}
\ee
Renormalizing the integration constant $C_2$, i.e., turning to $\tilde{C}_2= C_2/\ve$, we get the desired functional dependence for the flat space limit.

The other part of the (A)dS solution for the deformation function in eq. \rf{SolgenAdS} can be related to its flat space counterpart \rf{Sol1} as follows. For a real-valued $x$ we have
\be
\arccosh x=\ln \left(x+\sqrt{x^2-1} \right).
\la{arccoshkog}
\ee
In the flat space limit $L^2\ra \infty$ the proportional to $C_1$ part of \rf{SolgenAdS} turns out to be 
\be
\begin{split}
\arccosh \Bigg[-&\fr1{4\D\sin^2\th}\left(\fr{\D^2}{r^2_0}+4 r^2_0\sin^2\th-r^2_0\sin^2 2\th \right)\Bigg]\\
\equiv &\arccosh \Bigg[-\fr1{4\D\sin^2\th}\left(\fr{\D^2}{r^2_0}+r^2_0 {\cal F}(\th) \right)\Bigg].
\end{split}
\la{SolC1AdS20}
\ee
According to eq. \rf{arccoshkog}, the $\arccosh x$ series expansion looks like
\be
\arccosh x=\ln(2x)-\sum_{n=1}^\infty \left(\fr{(2n)!}{2^{2n}(n!)^2}\right) \fr{x^{-2n}}{2n},
\la{arccoshSer}
\ee
that, for the region far from the BH horizon gives
\be
\begin{split}
\arccosh \Bigg[-\fr1{4\D\sin^2\th}\left(\fr{\D^2}{r^2_0}+r^2_0 {\cal F}(\th) \right)\Bigg]\Bigg|_{r\gg 1}\approx \ln \left(-\fr{\D}{2 r^2_0 \sin^2 \th} \right)+{\cal O}(r)\\
=\ln\left(\fr{\D}{2 r^2_0 \sin^2 \th} \right)+i\pi+{\cal O}(r).
\end{split}
\la{C1expan}
\ee
Since the solution \rf{Sol1AdS} is determined up to an additional additive constant,  we can always fix this additional constant to recover the flat space solution \rf{Sol1} after rescaling the constant $C_1$.


\bsk\bsk

\end{document}